\documentclass[11pt]{article}
\pdfoutput=1

\usepackage[utf8]{inputenc}
\usepackage{multirow}
\usepackage{amsmath, amsfonts, amssymb}
\usepackage{comment}
\usepackage{graphicx,import}
\usepackage{psfrag}
\usepackage{amsthm}

\allowdisplaybreaks
\usepackage[dvipsnames,svgnames,table]{xcolor}
\usepackage{enumerate}
\usepackage{arydshln}
\usepackage{soul}
\usepackage{array}
 \usepackage{slashed}
 \usepackage{nicefrac}
 \usepackage[most]{tcolorbox}
\usepackage[font=small,labelfont=bf]{caption} 
\usepackage{subcaption}
\usepackage{mathtools}
\usepackage{colortbl}
\usepackage{mathrsfs}
 \usepackage{a4wide}
  \usepackage{tikz}
  \usepackage{tikz-cd}
  \usetikzlibrary{shapes.geometric}
  \usepackage{tcolorbox}
\usepackage{diagbox}

  \usepackage{color}
  \definecolor{dark-gray}{gray}{0.20}
  \definecolor{gray}{gray}{0.30}
  \definecolor{light-gray}{gray}{0.80}
  \definecolor{dark-red}{rgb}{0.7,0,0}
  \definecolor{dark-green}{rgb}{0.1,0.4,0}
  \definecolor{dark-blue}{rgb}{0.3,0.3,0.7}
  \definecolor{light-blue}{rgb}{0.8,0.8,1}
      \definecolor{swamp}{RGB}{240, 199, 197}
       \definecolor{landscape}{RGB}{180, 250, 199}
          \definecolor{undecided}{RGB}{252, 252, 197}
\definecolor{myRED}{HTML}{FF0000}
\definecolor{myGREEN}{HTML}{00AA00}
\definecolor{myBLUE}{HTML}{0055D4}
\definecolor{nORANGE}{RGB}{245, 154, 35}
\definecolor{nBLUE}{RGB}{0, 122, 255}
\definecolor{nGREEN}{RGB}{142, 196, 79}
\definecolor{nRED}{RGB}{216, 77, 36}
\definecolor{nPURPLE}{RGB}{115, 20, 205}
\definecolor{nPINK}{RGB}{197, 49, 140}

  \usepackage{pifont}
\usepackage{newunicodechar} 

\usepackage{setspace}

\newcommand{\beq}{\begin{equation}}  \newcommand{\eeq}{\end{equation}}
\newcommand{\bal}{\begin{aligned}}   \newcommand{\eal}{\end{aligned}}
\newcommand{\be}{\begin{equation}}
\newcommand{\ee}{\end{equation}}

\def\be{\begin{equation}}
\def\ee{\end{equation}}
\def\bea{\begin{eqnarray}}
\def\eea{\end{eqnarray}}

\newcommand{\dd}{\mathrm{d}}

\def\simleq{\; \raise0.3ex\hbox{$<$\kern-0.75em
      \raise-1.1ex\hbox{$\sim$}}\; }
   \def\simgeq{\; \raise0.3ex\hbox{$>$\kern-0.75em
      \raise-1.1ex\hbox{$\sim$}}\; }

\numberwithin{equation}{section}

\usepackage{jheppub}
\usepackage{hyperref}

\hypersetup{
	colorlinks=true,
	linkcolor=dark-blue,
	citecolor=dark-red,
	urlcolor=dark-green,
	linktoc=page
}

\theoremstyle{remark}

\title{\centering A missing link: \\ Brane networks and the Cobordism Conjecture}

\author{Markus Dierigl and Ignacio Ruiz}\affiliation{CERN, Theoretical Physics Department, 1211 Meyrin, Switzerland}
\emailAdd{markus.dierigl@cern.ch}
\emailAdd{ignacio.ruiz.garcia@cern.ch}

\preprint{CERN-TH-2026-108}

\abstract{The absence of global symmetries in a quantum gravity theory often requires the introduction of (new) symmetry-breaking defects, which appear as singular objects in the low-energy description. This has been formalized in the Cobordism Conjecture, which further relates the asymptotics of these defects to non-trivial deformation classes of the effective theory. In this work we investigate the symmetry-breaking defects for theories with a discrete symmetry $G$ encoded in the bordism groups $\Omega^{\xi}_2 (BG)$ and, in particular, its sub-class described in terms of the homology groups $H_2(BG;\mathbb{Z})$. Contrary to expectations we find that the defects are naturally described in terms of networks of codimension-two objects rather than isolated objects in codimension three. While in special situations linking configurations of defects are sufficient, our strategy generically predicts the existence of junctions, thus suggesting an extended applicability of the Cobordism Conjecture. We demonstrate the viability of this approach in four-dimensional supergravity theories originating from string and M-theory with a discrete Heisenberg group acting on its axionic degrees of freedom.}

\begin{document}

\hypersetup{pageanchor=false}
\makeatletter
\let\old@fpheader\@fpheader

\makeatother

\maketitle

\hypersetup{
    pdftitle={Understanding H2(BG,Z)},
    pdfauthor={Markus Dierigl, Ignacio Ruiz},
    pdfsubject={Understanding H2(BG,Z)}
}

\newcommand{\remove}[1]{\textcolor{red}{\sout{#1}}}

\section{Introduction}
\label{sec:intro}

Black holes and holographic theories provide strong evidence that global symmetries are absent in a theory of quantum gravity, see, e.g., \cite{Hawking:1975vcx,Zeldovich:1976vq,Zeldovich:1977be,Banks:1988yz,Kallosh:1995hi,Polchinski:1998rr,Banks:2010zn,Harlow:2018fse}. Thus, symmetries should be either gauged or broken below the quantum gravity scale. The interplay between symmetries and conserved charges offers another perspective; in the presence of a global symmetry the theory splits into various disconnected sectors labeled by their global charge. In this interpretation, gauging imposes consistency constraint (such as Gauss's law) that forbids certain sectors, while the breaking of a symmetry introduces connections between the formerly disconnected sectors. This logic has been made precise in the Cobordism Conjecture \cite{McNamara:2019rup}, which demands that quantum gravity theories (in $D$ dimensions) should not allow for any non-trivial deformation classes, mathematically described by bordism groups
\begin{equation}
    \Omega^{\text{QG}}_k = 0 \,, \quad 0 < k <D \,. 
\end{equation}
In particular, this requires that one can always deform the theory to nothing, and there can be no conserved symmetry charges.

While the gauging of global symmetries is an interesting subject in its own right, it typically leads to the introduction of other potentially dangerous conserved charges, see, e.g., \cite{Blumenhagen:2021nmi, Dierigl:2026sok}, and we will focus on symmetry breaking. Since we do not have the knowledge of the full structure of quantum gravity, we instead determine the non-trivial deformation classes of a low-energy effective theory described by the bordism groups $\Omega^{\xi}_k (BG)$, where $\xi$ refers to the required structures on the tangent bundle of spacetime and $BG$ allows for gauge fields of the symmetry group $G$.\footnote{These two pieces can also mix, leading to so-called twisted tangential structures, see \cite{Sati:2011rw, Pantev:2016nze, Tachikawa:2018njr, Freed:2019sco, Debray:2023yrs, Basile:2023knk, Debray:2024vso, Chakrabhavi:2025bfi} for some examples.} If these groups are non-trivial, we need to introduce new effects, beyond the low-energy effective theory, that trivialize them. Typically these are described by symmetry-breaking defects of codimension $(k+1)$, which have the non-trivial deformation class as asymptotic geometry. Indeed, this strategy has been successfully applied in various theories and produced various surprises, see, e.g., \cite{Dierigl:2022reg, Debray:2023yrs, Kaidi:2023tqo, Dierigl:2023jdp, Debray:2023rlx, Kaidi:2024cbx, Fukuda:2024pvu, Heckman:2025wqd, Torres:2026vxx, Anastasi:2026cus}.

\begin{figure}[hbt!]
    \centering
     \resizebox{0.33\textwidth}{!}{%
\begingroup%
  \makeatletter%
  \providecommand\color[2][]{%
    \errmessage{(Inkscape) Color is used for the text in Inkscape, but the package 'color.sty' is not loaded}%
    \renewcommand\color[2][]{}%
  }%
  \providecommand\transparent[1]{%
    \errmessage{(Inkscape) Transparency is used (non-zero) for the text in Inkscape, but the package 'transparent.sty' is not loaded}%
    \renewcommand\transparent[1]{}%
  }%
  \providecommand\rotatebox[2]{#2}%
  \newcommand*\fsize{\dimexpr\f@size pt\relax}%
  \newcommand*\lineheight[1]{\fontsize{\fsize}{#1\fsize}\selectfont}%
  \ifx\svgwidth\undefined%
    \setlength{\unitlength}{140.19039124bp}%
    \ifx\svgscale\undefined%
      \relax%
    \else%
      \setlength{\unitlength}{\unitlength * \real{\svgscale}}%
    \fi%
  \else%
    \setlength{\unitlength}{\svgwidth}%
  \fi%
  \global\let\svgwidth\undefined%
  \global\let\svgscale\undefined%
  \makeatother%
  \begin{picture}(1,0.76896093)%
    \lineheight{1}%
    \setlength\tabcolsep{0pt}%
    \put(0,0){\includegraphics[width=\unitlength,page=1]{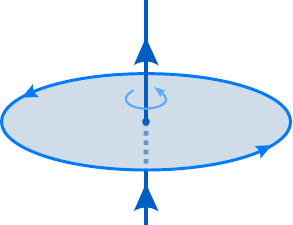}}%
    \put(0.02233422,0.52078651){\color[rgb]{0,0.47843137,1}\makebox(0,0)[lt]{\lineheight{1.25}\smash{\begin{tabular}[t]{l}$\color{nBLUE}{S^1_a}$\end{tabular}}}}%
    \put(0.51796401,0.6626976){\color[rgb]{0,0.47843137,1}\makebox(0,0)[lt]{\lineheight{1.25}\smash{\begin{tabular}[t]{l}$\color{nBLUE}{a}$\end{tabular}}}}%
  \end{picture}%
\endgroup%

    }
    \caption{Sketch of how a circle $S^1_a$ supporting a transition function given by $a\in G$ as one winds around it requires a codimension-two defect implementing such monodromy in order to be described as a boundary.}
    \label{fig:def1}
\end{figure}

In this work we investigate oriented and Spin manifolds with a discrete symmetry group $G$, which often appear in gravitational theories. These discrete symmetries can include dualities, large diffeomorphisms of internal spaces after compactification, as well as continuous symmetries which are broken to a discrete subgroup at low energies. The presence of such symmetries immediately produces viable candidates for extended objects in codimension-two, defined by the symmetry transformation $a \in G$ experienced when looping around the object, see Figure \ref{fig:def1}. However, for the absence of conserved charges only a subset of them is required, as captured by the associated bordism group in one dimension
\begin{equation}
    \Omega^{\xi}_1 (BG) = \Omega^{\xi}_1 (\text{pt}) \oplus H_1 (BG) = \Omega^{\xi}_1 (\text{pt}) \oplus \text{Ab}(G) \,.
\end{equation}
The first part describes purely gravitational conserved charges encoded in $\Omega_1^{\xi}(\text{pt})$ and is of less interest to us in this work. The charges induced by the non-trivial discrete symmetries are encoded in the first group homology $H_1(BG;\mathbb{Z})$ given by the Abelianization of $G$
\begin{equation}
    \text{Ab}(G) = \frac{G}{[G,G]} \,.
\end{equation}
As discussed for example in \cite{McNamara:2021cuo}, the reason for this reduction of necessary defects is that all elements that can be written as a product of commutators can be bounded by smooth, but topologically non-trivial, spacetime configurations, i.e., gravitational solitons and do not require the introduction of (new) singular objects.

\begin{figure}
    \centering
     \resizebox{0.8\textwidth}{!}{%
\begingroup%
  \makeatletter%
  \providecommand\color[2][]{%
    \errmessage{(Inkscape) Color is used for the text in Inkscape, but the package 'color.sty' is not loaded}%
    \renewcommand\color[2][]{}%
  }%
  \providecommand\transparent[1]{%
    \errmessage{(Inkscape) Transparency is used (non-zero) for the text in Inkscape, but the package 'transparent.sty' is not loaded}%
    \renewcommand\transparent[1]{}%
  }%
  \providecommand\rotatebox[2]{#2}%
  \newcommand*\fsize{\dimexpr\f@size pt\relax}%
  \newcommand*\lineheight[1]{\fontsize{\fsize}{#1\fsize}\selectfont}%
  \ifx\svgwidth\undefined%
    \setlength{\unitlength}{280.34250334bp}%
    \ifx\svgscale\undefined%
      \relax%
    \else%
      \setlength{\unitlength}{\unitlength * \real{\svgscale}}%
    \fi%
  \else%
    \setlength{\unitlength}{\svgwidth}%
  \fi%
  \global\let\svgwidth\undefined%
  \global\let\svgscale\undefined%
  \makeatother%
  \begin{picture}(1,0.41124189)%
    \lineheight{1}%
    \setlength\tabcolsep{0pt}%
    \put(0,0){\includegraphics[width=\unitlength,page=1]{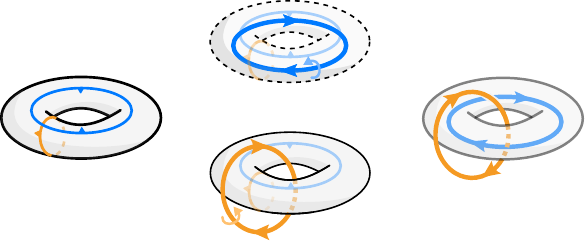}}%
    \put(0.14684073,0.28697204){\color[rgb]{0,0.47843137,1}\makebox(0,0)[lt]{\lineheight{1.25}\smash{\begin{tabular}[t]{l}$\color{nBLUE}{a}$\end{tabular}}}}%
    \put(0,0){\includegraphics[width=\unitlength,page=2]{Linking1.pdf}}%
    \put(0.49172217,0.23791211){\color[rgb]{0,0.47843137,1}\makebox(0,0)[lt]{\lineheight{1.25}\smash{\begin{tabular}[t]{l}$\color{nBLUE}{a}$\end{tabular}}}}%
    \put(0.06864873,0.10903796){\color[rgb]{0,0.47843137,1}\makebox(0,0)[lt]{\lineheight{1.25}\smash{\begin{tabular}[t]{l}$\color{nORANGE}{b}$\end{tabular}}}}%
    \put(0.34215627,0.02543487){\color[rgb]{0,0.47843137,1}\makebox(0,0)[lt]{\lineheight{1.25}\smash{\begin{tabular}[t]{l}$\color{nORANGE}{b}$\end{tabular}}}}%
  \end{picture}%
\endgroup%

    }
    \caption{Filling of the inside and outside of a 2-torus with transition functions $a$ and $b$, such that $[a,b]=1$. Embedded in $S^3\simeq\mathbb{R}^3\cup\{\infty\}$, the inclusion of two codimension-two defects implementing said monodromies is needed. Note that the two defects are linked, with total linking number 1.\vspace{0.15cm}\\    
    The arrows on a given $g\in G$ transition functions denote the direction of implementation (i.e., going in the arrow direction results in a $g$ transformation, while opposite to it implements $g^{-1}$). This translates to the orientation of defects and their monodromy through the right-hand-rule.}
    \label{fig:torusfilling}
\end{figure}

Continuing this strategy for higher-dimensional deformation classes $\Omega^{\xi}_k (BG)$ one expects to find further symmetry-breaking defects in codimension $(k+1)$. Here, we will focus on the next step, $k = 2$, for which one has
\begin{equation}
    0 \rightarrow \Omega^{\xi}_2(\text{pt}) \oplus H_1 \big(BG; \Omega^{\xi}_1 (\text{pt}) \big) \rightarrow \Omega^{\xi}_2 (BG) \rightarrow H_2 (BG) \rightarrow 0 \,,
    \label{eq:sesOmega2}
\end{equation}
which can be deduced from the associated Atiyah-Hirzebruch spectral sequence, see, e.g., \cite{Garcia-Etxebarria:2018ajm, Yonekura:2022reu}. We see that the second group homology $H_2(BG;\mathbb{Z})$ naturally appears. These non-trivial deformation classes can be described by $G$ gauge field backgrounds on genus-$g$ Riemann surfaces $\Sigma_g$, with $g \geq 1$. This poses a challenge for our understanding of the dimensionality of the symmetry-breaking defect, which would appear to be codimension-two rather than codimension-three, by filling the Riemann surface without metric singularity, see Figure \ref{fig:torusfilling}.\footnote{Note that the more singular configuration of a cone over $\Sigma_g$ with associated metric singularity would also be possible. We thank Jake McNamara for commenting on this.} Indeed this behavior already appeared in \cite{Braeger:2025kra} where the symmetry group SL$(3;\mathbb{Z})$ is perfect, i.e., its Abelianization is trivial, but codimension-two defects are revived through the non-triviality of $\Omega^{\text{Spin}}_2 (BG)$.

In the present work we will resolve this mismatch by identifying the obstruction to a deformation to nothing not as the individual monodromies, but their collective interplay. In particular, we see that the codimension-two defect does not appear isolated but has to pass through other $G$ transition functions, producing the codimension-three features expected. This also manifests on the level of the symmetry-breaking defects, which appear in non-trivial brane networks. For a very simple example consider an element of $H_2(BG;\mathbb{Z})$ which can be detected on a 2-torus, i.e., going around the two basis 1-cycles one encounters the monodromies $a$ and $b$, which necessarily have to commute. Filling the interior of the torus one needs to include a codimension-two $a$-defect winding around the non-trivial 1-cycle of the solid torus. However, equally well we could ``fill'' the outside by introducing a codimension-two $b$-defect passing ``through the hole'' of the torus. Doing both fillings at the same time we find a configuration of an $a$-defect linking a $b$-defect, which poses an obstruction to finding a pure gravitational spacetime without any defects. See Figure \ref{fig:torusfilling} for an illustration.

In general not all elements in $H_2(BG;\mathbb{Z})$ can be detected by a 2-torus and some require higher-genus Riemann surfaces. In these cases, while one still has the equivalence of filling the inside or outside by defects that link each other, one uncovers brane networks even as the symmetry-breaking defects themselves. Moreover, while in special situations a linking configuration of codimension-two objects can be enough to trivialize $\Omega^{\xi}_2 (BG)$, generically one has to include brane junctions.\footnote{We refer the reader to the coordinates \texttt{46°12'04.4"N 6°07'17.8"E} for a visualization of such junction.} Thus, we can use the properties of the symmetry group $G$ together with the Cobordism Conjecture to not only predict the existence of potentially new objects, but also suggest their configurations and networks.

The manuscript is organized as follows: In Section~\ref{sec:Hopf} we describe how (part of) the second bordism group $\Omega^{\xi}_2 (BG)$ is described in terms of the second group homology $H_2(BG;\mathbb{Z})$ and how these are described via non-trivial $G$-bundles on genus-$g$ Riemann manifolds, via Hopf's theorem. In Section~\ref{sec:link} we use this description to explore the associated symmetry-breaking defects predicted by the Cobordism Conjecture. In particular, we discuss the importance of brane networks including the linking and intersection (junctions) of codimension-two objects. We apply this general strategy to type IIA compactifications on Calabi-Yau 3-folds in Section~\ref{sec:juncIIA} involving Heisenberg groups acting on the axionic degrees of freedom. After concluding and pointing out future research directions in Section~\ref{sec:concl}, we generalize the string theory examples in Appendix~\ref{app:Mth}.

\section{Hopf's theorem and the Cobordism Conjecture}
\label{sec:Hopf}

In this work we want to study the defects breaking global symmetries associated to a non-trivial bordism group $\Omega^{\xi}_2 (BG)$, for discrete groups $G$ and tangential structure $\xi$. Since we do not consider twisted tangential structures these groups split into a pure gravitational part and a part that receives gauge contributions described by the reduced bordism group $\widetilde{\Omega}^{\xi}_2 (BG)$
\begin{equation}
    \Omega^{\xi}_2 (BG) = \Omega^{\xi}_2 (\text{pt}) \oplus \widetilde{\Omega}^{\xi}_2 (BG) \,.
\end{equation}
In the following we will mainly focus on $\xi$ denoting orientation (SO) or Spin structure. 
\begin{figure}
\centering
\begin{tikzpicture}[scale=0.7]
    \draw[->] (0,0) -- (12.2,0) node[right] {$p$};
    \draw[->] (0,0) -- (0,3.2) node[above] {$q$};
    \draw[xstep=4cm,gray!20] (0,0) grid (12,3);
    \foreach \x in {0,1,2}
        \node[below] at (4*\x+1, 0) {\small $\x$};
    \foreach \y in {0,1,2}
        \node[left] at (0,\y+0.5) {\small $\y$};
    \node at (2, 0.5) {$\mathbb{Z}$};
    \node at (6, 0.5) {$H_1(BG;\mathbb{Z})$};
    \node at (10, 0.5) {$H_2(BG;\mathbb{Z})$};
    \node at (2, 1.5) {$\Omega^{\xi}_1 (\text{pt})$};
    \node at (2, 2.5) {$\Omega^{\xi}_2 (\text{pt})$};
    \node at (6, 1.5) {$H_1\big( BG; \Omega^{\xi}_1 (\text{pt}) \big)$};
    \node at (6, 2.5) {$*$};
    \node at (10, 1.5) {$*$};
    \node at (10, 2.5) {$*$};
\end{tikzpicture}
\caption{$E_{p,q}^2$ for the Atiyah-Hirzebruch spectral sequence for $\Omega^{\xi}_{\bullet}(BG)$.}
\label{fig:AHSS}
\end{figure}
Considering the second page of the Atiyah-Hirzebruch spectral sequence $E^2_{p,q}$ depicted in Figure~\ref{fig:AHSS}, and noting that for degree reasons the differentials are all trivial we find
\begin{equation}
0 \rightarrow H_1 \big( BG; \Omega^{\xi}_1 (\text{pt}) \big) \rightarrow \widetilde{\Omega}^{\xi}_2 (BG) \rightarrow H_2 (BG;\mathbb{Z}) \rightarrow 0 \,.
\end{equation}
Since $\Omega^{\text{SO}}_1 (\text{pt})$ is trivial we find that in two dimensions the oriented bordism class is fully determined by group homology, while for Spin bordism one has a potential extension by $H_1(BG;\mathbb{Z}_2)$. In both cases $H_2 (BG;\mathbb{Z})$ plays a central role.\footnote{Even in the case of Spin bordism it is enough to trivialize classes originating from $H_2(BG;\mathbb{Z})$, since if the extension is non-trivial they will also trivialize the classes coming from $H_1$. If the extension is trivial on the other hand these elements already appear for $\Omega^{\text{Spin}}_1 (BG)$ and are trivialized by the introduction of codimension-two defects there.} To understand the non-trivial classes in $\Omega^{\xi}_2 (BG)$ it is therefore crucial to have a good understanding of the elements in $H_2 (BG;\mathbb{Z})$ as well as their physical interpretation. 

The obstruction of non-trivial classes originating from $H_2(BG;\mathbb{Z})$ is entirely encoded in the gauge sector. For discrete $G$ and spacetime manifold $X$ this is captured by the configurations of the associated gauge field, whose different topological sectors are given by homotopy classes of  (see, e.g., \cite{Dijkgraaf:1989pz, Freed:1991bn, Hatcher:478079})
\begin{equation}
    [X,BG] \simeq [X, K(G,1)] \,,
\end{equation}
where we used that $BG$ is (homotopy) equivalent to an Eilenberg-MacLane space $K(G,1)$. For Abelian groups $G$, this can be identified with $H^1(X;G)$, while for non-Abelian groups it can be understood as
\begin{equation}
    [X,BG] = \frac{\text{Hom} \big( \pi_1(X); G \big)}{G} \,.
\end{equation}
This means that to each (homotopically) non-trivial closed path in $X$ we associate a transition function given in terms of a group element $a \in G$, up to overall gauge transformations. Since the map is a group homomorphism these transition function have to respect consistency conditions encoded in the group properties of $\pi_1(X)$, the fundamental group of $X$.

Let us focus on closed two-dimensional spacetimes $X$, which for oriented and Spin manifolds are given by Riemann surfaces $\Sigma_g$. Since for the 2-sphere $S^2$ we have that $\pi_1(S^2) = 0$, there are no non-trivial gauge bundles underlining that the defects are not given in terms of codimension-three defects with $S^2$ as the asymptotic space. Next, we can describe gauge fields on the 2-torus $T^2$, whose fundamental group is $\mathbb{Z} \oplus \mathbb{Z}$. Gauge configurations are hence classified by two commuting elements $a$ and $b$ in $G$, which account for the transition functions around the basis of $\pi_1(T^2)$, up to overall gauge transformations
\begin{equation}
    [a,b] = 1 \,.
\end{equation}
Finally, for genus $g \geq 2$ Riemann surfaces the fundamental group is generated by $2g$ elements with associated transition functions determined by the group elements $\{a_1, b_1 , \dots a_g, b_g\}$, which satisfy the condition
\begin{equation}
    [a_1,b_1] \dots [a_g,b_g] = 1 \,.
    \label{eq:Gconscondition}
\end{equation}
With this description of $G$ bundles on $\Sigma_g$ we can move to a description of $H_2(BG;\mathbb{Z})$ in terms of a presentation of the group $G$.

\subsection{Bordism interpretation of Hopf's theorem}

In the following we will focus on discrete, but not necessarily finite, groups $G$, that often appear in (super)gravity theories. Such discrete groups can be described in terms of a presentation (e.g., \cite{BrownCohomologyGroups})
\begin{equation}
    G = \frac{F}{R} \,,
\end{equation}
where $F$ is a free group with generators $\{ f_i \}$ and $R$ is the normal closure of the set of relations, forming a normal subgroup of $F$. We denote the group as follows\footnote{We only consider groups with a finite number of generators $f_j$.}
\begin{equation}
    G = \langle f_1 \,, \dots \,, f_n | \{ r_i \} \rangle \,,
\end{equation}
given in terms of the generators $f_j$ and relations $r_i$. 

Let us describe some simple examples. The cyclic group $\mathbb{Z}_n$ is given in terms of a single generator $x$ subject to one relation
\begin{equation}
    \mathbb{Z}_n = \langle x | x^n \rangle \,.
\end{equation}
The dihedral group of order $2n$ can be described as
\begin{equation}
    D_{2n} = \langle x, r | x^n \,, r^2 \,, r x r x \rangle \,,
\end{equation}
where, interpreting $D_{2n}$ as the symmetry group of an $n$-gon, $x$ is rotation by $2\pi/n$ and $r$ is reflection along a symmetry axis of the associated $n$-gon. We can also describe the surface groups $\pi_1(\Sigma_g)$ in this way. As mentioned above they are given in terms of $2g$ generators $\{ a_i, b_i\}$ obeying a single relation
\begin{equation}\label{eq.surface group}
    \pi_1(\Sigma_g) = \langle a_1, b_1, \dots, a_g, b_g | [a_1,b_1]  \dots [a_g,b_g] \rangle \,,
\end{equation}
which asserts that this combination of paths can be contracted to a point on $\Sigma_g$ and hence is null-homotopic.

The information of the presentation of the discrete group also allows to determine the second homology group using Hopf's theorem in group homology, see, e.g.,
\cite{BrownCohomologyGroups}
\begin{equation}   \label{eq. hopf}
H_2(BG;\mathbb{Z})\simeq \frac{R\cap[F,F]}{[R,F]} \;.
\end{equation}
The numerator states that elements in $H_2(BG;\mathbb{Z})$ can be expressed as a product of commutators of the free generators in $F$ which is part of the relations $R$. This means that an element of the second homology can associated to
\begin{equation}
    [f_i, f_j] \dots [f_k,f_l] = 1 \in G \,.
    \label{eq:Hopfnumerator}
\end{equation}
Comparing to the discussion of $G$ bundles on $\Sigma_g$ above this implies that each non-trivial element of $H_2(BG;\mathbb{Z})$ can be realized as a gauge configuration on $\Sigma_g$ with $g$ bigger or equal the number of commutators in \eqref{eq:Hopfnumerator}. The denominator states that (products of) commutators with relations as one of its entries are trivial.

Let us use the formula \eqref{eq. hopf} in the examples discussed above. For the cyclic group we have only a single generator and thus $[F,F]$ is trivial, which correctly implies that
\begin{equation}
    H_2 (B\mathbb{Z}_n;\mathbb{Z}) = 0 \,.
\end{equation}
For the dihedral group $D_{2n}$ and $n$ even one can find a relation given by
\begin{equation}
    r x^{n/2} r x^{n/2} = r x^{n/2} r^{-1} x^{-n/2} = [r,x^{n/2}] \,,
\end{equation}
which is given by a commutator without any relation, which suggests that $H_2 (BD_{2n};\mathbb{Z})$ is non-trivial. However, we can rewrite the square of this commutator as follows
\begin{equation}
    [r, x^{n/2}][r, x^{n/2}] = [r,x^n] \,,
\end{equation}
showing that it is part of $[R,F]$ and hence trivial in $H_2(BG;\mathbb{Z})$. For $n$ odd no such element can be found and we conclude
\begin{equation}
    H_2 (BD_{2n};\mathbb{Z}) = \begin{cases} \mathbb{Z}_2 \,, \quad n \text{ even} \,, \\
    0 \,, \quad n \text{ odd} \end{cases} \,,
\end{equation}
reproducing the correct result, see, e.g., \cite{HandelD2m}. Finally, for the surface group $\pi_1(\Sigma_g)$, $g \geq 1$, the relation is already given in terms of a product of commutators and no power of it vanishes leading to
\begin{equation}\label{eq.surfacegroupH2}
    H_2 \big( B \pi_1 (\Sigma_g); \mathbb{Z} \big) = \mathbb{Z} \,, \quad g \geq 1 \,, 
\end{equation}
which can also be deduced from the realization that $B\pi_1(\Sigma_g) \simeq \Sigma_g$.

\begin{figure}
    \centering
     \resizebox{0.63\textwidth}{!}{%
    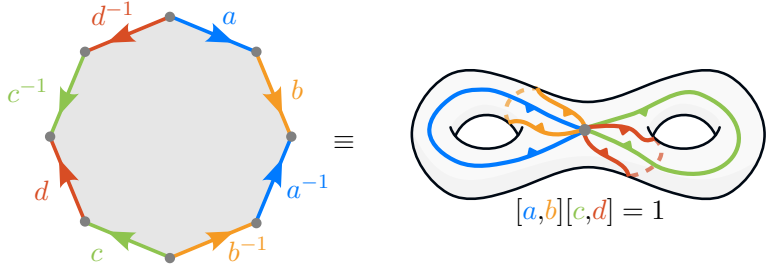
    }
    \caption{Sketch of the construction of a genus $g=2$ surface $\Sigma_2$ from the relation $[a,b][c,d]=aba^{-1}b^{-1}cdc^{-1}d^{-1}=1$, where the non-trivial 1-cycles are interpreted as non-relation elements from the group $G$. }
    \label{fig:8gon}
\end{figure}

We therefore find that elements $H_2(BG;\mathbb{Z})$ that necessarily survive as elements in $\Omega^{\xi}_2(BG)$ are given by genus-$g$ Riemann surfaces with a non-trivial $G$ bundle classified by the transition functions around the non-trivial closed paths. These further need to satisfy the relation of the surface group $\pi_1(\Sigma_g)$, see Figure \ref{fig:8gon}. Before we explore which defects are needed to describe such manifolds with $G$ bundle as the boundary of a three-dimensional manifold, we discuss what happens in case, despite the bundle on $\Sigma_g$ being non-trivial, the configuration describes a trivial element in $H_2 (BG;\mathbb{Z})$.

\subsection{Bordism to nothing}

To explore deformations to nothing imagine a Riemann surface $\Sigma_g$, with $g \geq 1$, and non-trivial transition functions on each basis element of $\pi_1(\Sigma_g)$ satisfying the consistency condition \eqref{eq:Gconscondition}. This configuration might still be trivial in $\Omega^{\text{SO}}_2 (BG)$, where for simplicity we focus on oriented bordism so we do not have to keep track of the Spin structure (which requires pinching cycles to support anti-periodic boundary conditions for the fermions). If this is the case it means that it describes the trivial element in $H_2(BG;\mathbb{Z})$ and by $\eqref{eq. hopf}$ is contained in $[R,F]$. In this section we want to explore, how one can construct a bounding 3-manifold, see also \cite{PutmanHopf}.

\begin{figure}[hbt!]
    \centering
     \resizebox{\textwidth}{!}{%
    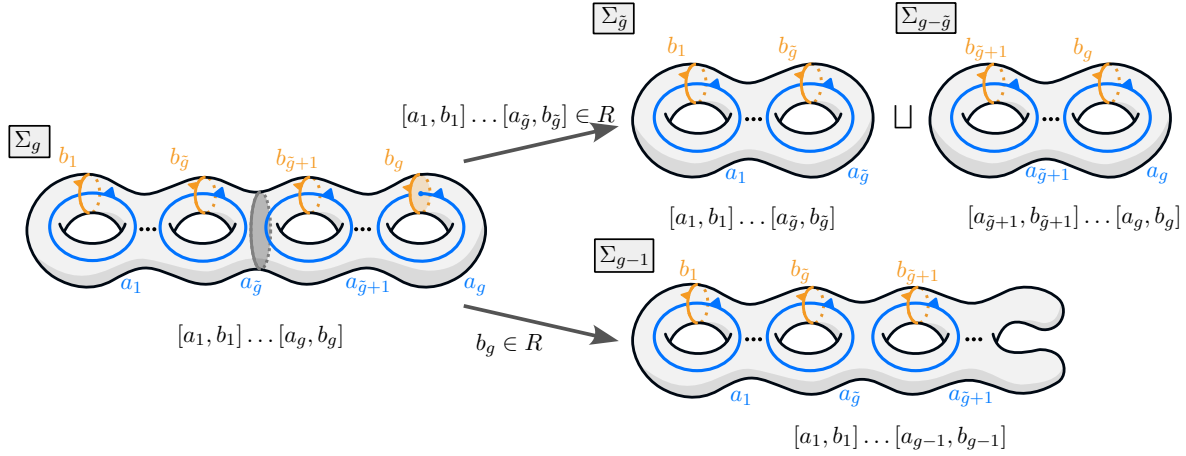
    }
    \caption{Sketch of how a topology change on a $[\Sigma_g]\in H_2(BG;\mathbb{Z})$ representative can occur when certain transition function along a 1-cycle is trivial. We depict the two possible cases, i.e., when a set of commutators is in $R$, which allows the splitting of $\Sigma_g$ into different connected components (with same total genus), or when a given monodromy \emph{inside} a commutator is in $R$, which allows the pinching of a 1-cycle, and the lowering of the genus. For the sake of simplicity we only draw the transition functions associated to the generators of $\pi_1(\Sigma_g)$, without depicting the additional $[a_i,b_i]$ lines.}
    \label{fig.split}
\end{figure}

First we note that if one of the transition functions is described by a relation, i.e., trivial in $G$ one can pinch $\Sigma_g$ and transition smoothly to $\Sigma_{g-1}$ or split the Riemann surface into two disconnected components $\Sigma_{\tilde{g}} \sqcup \Sigma_{g-\tilde{g}}$, see Figure~\ref{fig.split}. If, the original configuration is already in the form $[R,F]$ on can perform this operation several times to end up with the disconnected sum of spheres, which necessarily describe the trivial element in $H_2(BG;\mathbb{Z})$. In general, one first needs to execute deformations to reach this form.

\begin{figure}[hbt!]
    \centering
     \resizebox{\textwidth}{!}{%
    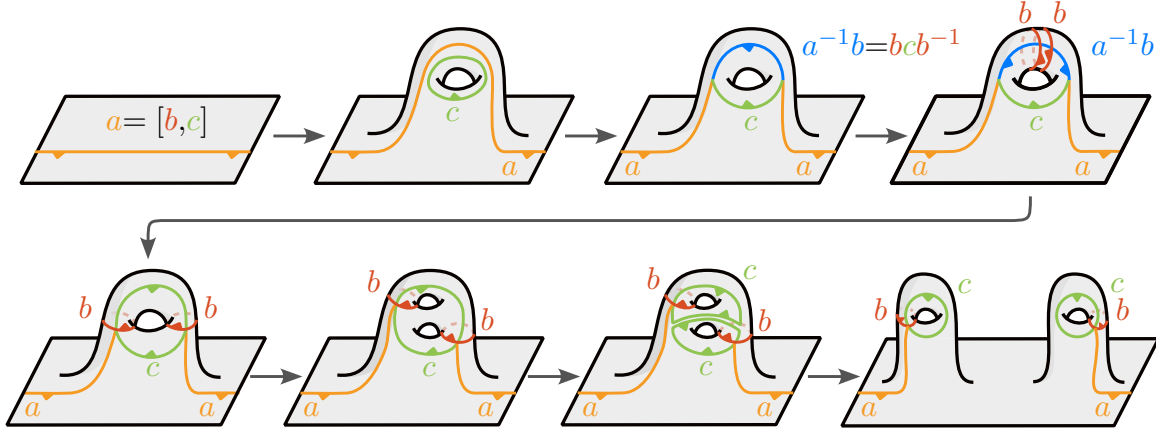
    }
    \caption{Sketch of how a transition function written as a commutator $a=[b,c]$ can be broken via the nucleation of two gravitational solitons. Note that after the above process, the 1-cycle dual to that previously supporting the transition function $a$ carries trivial monodromy, and thus can be pinched.}
    \label{fig:break_line}
\end{figure}

For that it will be very useful to study, how one can ``split open'' transition functions by increasing the topological complexity of the spacetime manifold. In particular, assume that one transition function is described by $a \in G$, which can be written as a commutator\footnote{Similar constructions also work for products of commutators in which case one would generate higher-genus configurations.}
\begin{equation}
    a = [b,c] \,.
    \label{eq:addhandle}
\end{equation}
By the nucleation of a pair of gravitational solitons of genus one, see Figure~\ref{fig:break_line}, one can then split the transition function. Note that the same gravitational solitons appear in the discussion of $\Omega^{\xi}_1(BG)$ as a smooth manifold bounding a circle twisted with transition function $a$. Implementing this operation for one of the transition functions on $\Sigma_g$, one obtains a new Riemann surface $\Sigma_{g+2}$ where one non-trivial closed path, that crossed the now split transition function, carries trivial monodromy and therefore can be pinched. The result is $\Sigma_{g+1}$, where the handle with transition function given by $a$ is replaced by two handles with transition function given by $b$ and $c$ satisfying \eqref{eq:addhandle}. With these and similar transitions described in \cite{PutmanHopf} one can always reach a configuration given by a Riemann surface with transition functions in $[R,F]$, which then can be deformed to nothing.

\subsection{Examples}

Let us illustrate this strategy in some examples.

For the dihedral group with $n$ even we had $H_2 (BD_{2n};\mathbb{Z})$ given by $\mathbb{Z}_2$, with the associated element given in terms of a single commutator
\begin{equation}
    [r,x^{n/2}] = 1 \,.
\end{equation}
From this we learn that $\Omega^{\text{SO}}_2 (BD_{2n})$ is generated by a 2-torus $T^2$ with transition functions $r$ and $x^{n/2}$, both elements of order two, around the basis of 1-cycles. In this case these can be directly related to the non-trivial elements of $\Omega^{\text{SO}}_1 (BD_{2n})$ given by the Abelianization of $D_{2n}$ 
\begin{equation}
    \Omega^{\text{SO}}_1 (BD_{2n}) = \text{Ab}(D_{2n}) = \mathbb{Z}_2 \oplus \mathbb{Z}_2 \,, \quad n \text{ even} \,,
\end{equation}
which is precisely generated by the two elements $r$ and $x^{n/2}$ above. Even for Spin structure, there is no non-trivial extension and the generator above corresponds to a generator of a $\mathbb{Z}_2$ subgroup, see, e.g., \cite{bookBrunerGreenlees, Davighi:2022icj}.

The story changes for the group SL$(n;\mathbb{Z})$ with $n \geq 5$, which has a presentation in terms of Steinberg relations \cite{Milnor_AlgebraicKTHeory}. Importantly these groups are perfect, which means
\begin{equation}
    \Omega^{\xi}_1 \big(B \mathrm{SL}(n;\mathbb{Z}) \big) \supset H_1 \big(B \mathrm{SL}(n;\mathbb{Z}); \mathbb{Z} \big) \simeq \text{Ab} \big(\mathrm{SL}(n;\mathbb{Z}) \big) = 0  \,, \quad n \geq 3 \,.
\end{equation}
This means that the Cobordism Conjecture would not predict any defects in codimension-two. However, one finds that
\begin{equation}
    H_2 \big( B \mathrm{SL}(n;\mathbb{Z}); \mathbb{Z} \big) = \mathbb{Z}_2 \,, \quad n\geq5
\end{equation}
The associated element can be written as a single commutator. 
In particular, we can define the SL$(n;\mathbb{Z})$ elements, see, e.g., the discussion in \cite[Section 9 to 11]{Milnor_AlgebraicKTHeory},
\begin{equation}
    h_{12} = \text{diag}(-1,-1,1,1,\dots, 1) \,, \quad h_{13} = \text{diag}(-1,1,-1,1,\dots, 1) \,.
\end{equation}
These matrices commute and their commutator corresponds to the non-trivial element of $H_2 \big(B\mathrm{SL}(n;\mathbb{Z});\mathbb{Z}\big)$ with generator given by $T^2$ with the associated transition functions. In the case SL$(n;\mathbb{Z})$ with $n \in \{3\,, 4\}$, one finds $H_2\big(B\mathrm{SL}(n;\mathbb{Z});\mathbb{Z}\big) = \mathbb{Z}_2 \oplus \mathbb{Z}_2$ and a particular basis for the monodromies for $n=3$ can be found in \cite{SOULE19781, Braeger:2025kra}, see also \cite{986c62b68cc94b76bcbc5676fac2e603}. Since these groups are perfect, the reduced Spin bordism groups $\widetilde{\Omega}^{\text{Spin}}_2 \big( B \mathrm{SL}(n;\mathbb{Z})\big)$ are identical to $H_2 \big(B\mathrm{SL}(n;\mathbb{Z});\mathbb{Z}\big)$ and generated by the same manifolds as above.

With this we can also sharpen the argument of \cite{Braeger:2025kra} stating that gravitational solitons, despite being able to bound a circle with any SL$(n;\mathbb{Z})$ transition function, cannot be used to trivialize the 2-torus above. First, let us write one of the transition functions, say $h_{12}$, as a product of commutators which always works since the group is perfect\footnote{The minimal number of commutators an element of the commutator group, $x\in [G,G]$ can be expressed as is known as \emph{commutator length} of $x$, ${\rm cl}(x)$. While we will not use this concept along the paper, we refer to \cite{lasHeras:2026xdl} for its role in the context of the Cobordism Conjecture, and the problems it might entail in groups $G$ for which ${\rm cl}(x)$ is not bounded for $x\in G$.}
\begin{equation}
    h_{12} = [a_1,b_1] \dots [a_k,b_k] \,.
\end{equation}
We can then plug this into the expression defining the non-trivial element in $H_2(B\mathrm{SL}\big(n;\mathbb{Z});\mathbb{Z}\big)$
\begin{equation}
    1 = [h_{12},h_{13}] = \big[ [a_1, b_1] \dots [a_k,b_k] , h_{13}\big] \,.
\end{equation}
Using Leibniz rule, $[ab,c]= a[b,c]a^{-1}[a,c]$, several times, we can reduce the expression to a product of (conjugates) of the form
\begin{equation}
    \big[[a_j , b_j], h_{13} \big] = \big[b_j, [a_j,h_{13}]\big] \big[ a_j, [b_j, h_{13}] \big] \,,
\end{equation}
where we used the Jacobi identity. Since we started with a non-trivial element in homology, by Hopf's formula we find that not all $[a_j,h_{13}]$ and $[b_j,h_{13}]$ can be trivial, since otherwise the resulting element would be in $[F,R]$. However, this would be required for the gravitational soliton to glue appropriately across the $h_{13}$ transition function, see \cite[Figure 16]{Braeger:2025kra}. Thus, these classes indeed require the introduction of (singular) defects.

Finally one might ask the question whether all elements in $H_2 (BG;\mathbb{Z})$ can be associated to a single commutator. The easiest way to see that this is not the case is the surface group $\pi_1 (\Sigma_g)$ itself. It has a single relation given in terms of the product of $g$ commutators and cannot be reduced any further, thus, requiring the description in terms of a genus $g$ surface. We will also discuss the Heisenberg groups $\mathrm{H}_{2k +3}(\mathbb{Z})$ which require $g =2$ in a string theory example in Section~\ref{sec:juncIIA}.


\section{Linking and junctions}
\label{sec:link}

Once we have identified a subset of non-trivial classes in $\Omega^\xi_2(BG)$ described by Riemann surfaces with a non-trivial $G$ bundle, we switch our attention to the defects needed to trivialize them. While naively one would expect a genuine codimension-three defect playing such a role \cite[Section 3.1]{McNamara:2019rup}, since $BG$ is aspherical, such defect cannot support a non-trivial $G$ configuration around it\footnote{Note that in the case of unoriented manifolds there is one free quotient of the 2-sphere given by $\mathbb{RP}^2 \simeq S^2/\mathbb{Z}_2$ which would also be a viable asymptotic for a genuine codimension-three defect. Since this manifold has $\pi_1(\mathbb{RP}^2) = \mathbb{Z}_2$ it can also support non-trivial discrete gauge fields, but only of very special type described by Hom$\big( \mathbb{Z}_2; G\big)$.} without metric singularities, and instead codimension-two defects will naturally appear. These are the same type one typically  would have associated to classes in $\Omega_1^{\xi}(BG)$. In this Section we will resolve this conflict by noting that these codimension-two objects appear wrapped on closed curves and are generically involved in non-trivial brane configurations which capture the codimension-three nature. The importance of wrapping the symmetry defects on non-trivial cycles of the bordism generator has also been stressed in \cite[Section 3.3]{Ruiz:2024jiz}.

\subsection{Genus one}

For a non-trivial element of $H_2(BG;\mathbb{Z})$ to be described by a genus-one surface, i.e., a 2-torus $T^2$, Hopf's formula \eqref{eq. hopf} demands that the associated element $R \cap [F,F]$ can be expressed as a single commutator. This means that the configuration is determined by two commuting transition functions $a,b \in G$, $[a,b]=1$ (where of course both $a$ and $b$ need to be non-trivial elements).

As discussed in \cite{Braeger:2025kra}, one choice of defect that allows us to ``fill'' the torus and trivialize the $\Omega_2^\xi(BG)$ class is given by a codimension-two defect implementing the monodromy $a$, located in the ``interior'' of $T^2$. In this way it implements to correct transition function on the $T^2$. Without a second defect, the $b$ transition function must be continued to the interior of the solid torus and crosses the defect, however, since the two transitions commute, $[a,b]=1$, the defect is not altered in the process, can close, and the full configuration is well-defined. The solid torus given by filling the $S^1$ with $a$ transition function by adding the interior defect then trivializes the associated bordism class, see Figure~\ref{fig:torusfilling}.\footnote{From the Morse-Bott theoretical interpretation in \cite{Ruiz:2024jiz}, the bordism manifold corresponds to $\mathcal{B}_3=S^1\times D^2$, with the critical locus of the Morse-Bott function (i.e., the location of the bordism defect) being $S^1$.} 

Alternatively, through identical arguments we could have chosen to fill the ``outside'' of the torus by considering a codimension-two defect implementing a monodromy $b$. Indeed, the (smooth)\footnote{Allowing for non-smooth embedding might result in pathological situations like Alexander's horned sphere \cite{AlexanderHORNED}, where the interior of $S^3\backslash{}S^2$ is simply connected, but the exterior is not.} embedding of $T^2$ in $S^3\simeq \mathbb{R}^3\cup\{\infty\}$ splits $S^3$ in two different connected components with topology $S^1\times D^2$, and thus deformation-retracts to $S^1$. Thus, after filling both the inside and the outside of the $T^2$ embedded in $S^3$ one ends up with two (positively) linking codimension-two defects with monodromy given by $a$ and $b$, respectively, see Figure~\ref{fig:torusfilling}. We therefore see that the defect is not only a compactified codimension-two defect, but crucially contains a non-trivial transition function that implements a linking. This explains the mismatch in the dimensionality of the bordism groups for genus-one surfaces.

\subsection{Higher-genus and linking conditions
\label{ss.HigherGenusAndLinking}}

After describing the simple genus-one case (i.e., where the non-trivial class in $H_2(BG;\mathbb{Z})$ can be represented by a single commutator) we can consider the case with $g \geq 2$, associated to 
\begin{equation}
    [a_1,b_1]\dots[a_g,b_g]\in \frac{R\cap[F,F]}{[R,F]} \,,
\end{equation}
and represented by a genus-$g$ surface $\Sigma_g$ with the appropriate $G$ bundle. We can then take a (smooth) embedding $\Sigma_g\subset S^3$, corresponding to the genus-$g$ Heegaard splitting of the 3-sphere \cite{WALDHAUSEN1968195}. The two components $V_1\sqcup_{\Sigma_g}(-V_2)=S^3$ (where $-V_2$ denotes $V_2$ with the orientation flipped), with $\partial V_1=\partial V_2=\Sigma_g$, correspond to solid handlebodies of genus $g$, and have homology
\begin{equation}
    H_\bullet(V_1;\mathbb{Z})\simeq H_\bullet(V_2;\mathbb{Z})\simeq(\mathbb{Z},\mathbb{Z}^g,0,0,\dots)\,.
\end{equation}
This follows from the observation that they deformation retract to the wedge sum $\bigvee_{i=1}^g S^1$, the bouquet of $g$ circles joining at a point. One could then try to find the defect configuration allowing the filling of the interior of $V_1$ and $V_2$ in the following way.

\begin{figure}[htb ]
    \centering
    \resizebox{0.95\textwidth}{!}{%
    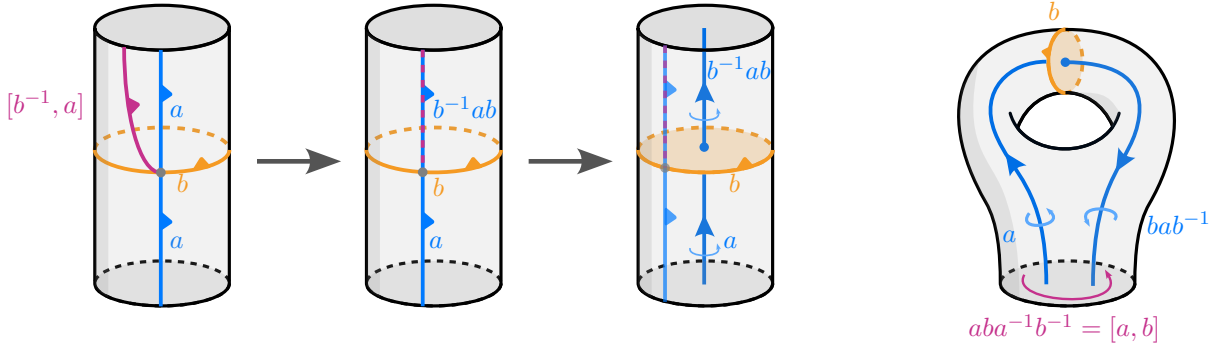
    }
    \caption{Illustration of the transformation $a\to b^{-1}ab$ after a defect with monodromy $a$ crosses a transition function $b$. On the right, we see how, upon ``closing'' the associated handle where the defect $a$ cross the transition function $b$ with opposite orientation, the total monodromy is given by $a(bab^{-1})^{-1}=[a,b]$}
    \label{fig.transDEF}
\end{figure}

\begin{figure}[htb]
    \centering    \resizebox{\textwidth}{!}{%
    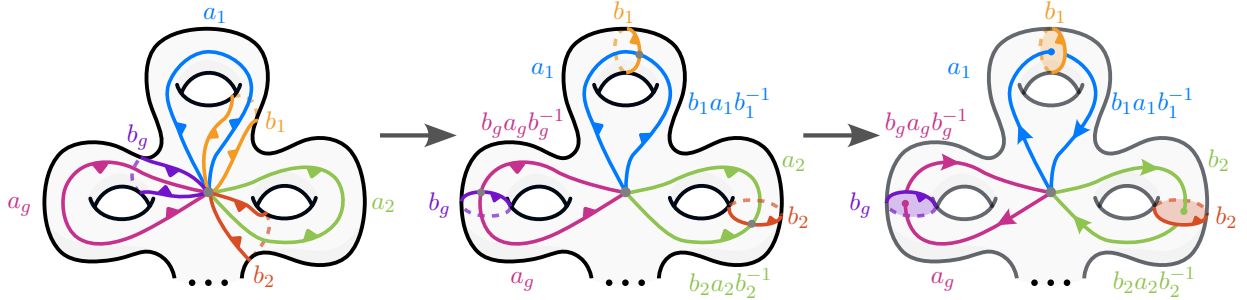
    }
    \caption{Sketch of the arrangement of transition functions for a genus-$g$ Riemann surface associated to a non-trivial element in $H_2(BG;\mathbb{Z})$ represented by $[a_1,b_1]\dots[a_g,b_g]$. In the center, we show how the $b_i$ transition function can be deformed to have an orientation opposite to $a_i$. Finally, on the right we depict a bouquet of $\{a_i\}_{i=1}^{g}$ defects allowing the filling of one component of $S^3\backslash\Sigma_g$, each of which goes through the $b_i$ transition function (with opposite orientation).}
    \label{fig:defectsGenusgtr2}
\end{figure}

Without loss of generality, we can ``fill'' the 1-cycles supporting the $a_i$ transition functions by the inclusion of a bouquet of codimension-two defects implementing the appropriate monodromy. Since $[a_i,b_i]\neq 1$ passing through the $b_i$ transition function implements a change of the monodromy of the defect \cite{Delgado:2024skw} (see also Figure \ref{fig.transDEF})
\begin{equation}
    a_i\to b_i^{-1} a_ib_i \,,
\end{equation}
in such a way that going around the handle supporting $\{a_i,b_i\}$ implements the monodromy $a_i(b_i^{-1} a_ib_i)^{-1}=[a_i,b_i]$. The joining of the the $g$ defects at the intersection point of the bouquet (notably a codimension-three point) implements the consistency condition $[a_1,b_1]\dots[a_g,b_g]=1$. This is illustrated in Figure~\ref{fig:defectsGenusgtr2}. Analogously, one could have ``filled'' the exterior component with $b_i$ defects intersecting the $a_i$ transition functions.

Unlike in the genus-one case, the arrangement of defects depicted in Figure~\ref{fig:defectsGenusgtr2} is highly singular, since it involves the \emph{junction} between different defects, a non-trivial property that might not be guaranteed without knowing the microscopic details of and UV realization of the defects, see \cite{Strominger:1995ac,Hanany:1996ie,Sen:1997xi,Aharony:1997bh,DeWolfe:1998zf} for some well known examples in the literature. However, it might very well be the case that no defect junctions are needed at all and, inspired by the linking of defects from Figure~\ref{fig:torusfilling}, the bouquet of defects can be resolved to a non-intersecting configuration given by a linking arrangement (albeit now in the same component of $S^3\backslash\Sigma_g$), see, e.g., Table~\ref{tab:linkingArrangements}.

To explore this possibility, we first take the crossing of two defects, with monodromies $\color{nBLUE}{a}$ and $\color{nORANGE}{b}$.  Depending on the sign of the crossing, i.e., whether $\color{nBLUE}{a}$ crossed above or below $\color{nORANGE}{b}$, the monodromy of the defect changes as 
\begin{equation}
\begin{array}{ccc}
   \begin{tikzcd}
\color{nORANGE}{aba^{-1}}\arrow[from=dr,nORANGE]  & \color{nBLUE}{a} \arrow[from=dl, crossing over,nBLUE] \\
\color{nBLUE}{a} & \color{nORANGE}{b}
\end{tikzcd}  &\qquad& \begin{tikzcd}
\color{nORANGE}{b}  & \color{nBLUE}{b^{-1}ab} \arrow[from=dl,nBLUE] \\
\color{nBLUE}{a}& \color{nORANGE}{b}\arrow[ul,crossing over,nORANGE]
\end{tikzcd} \\
  ~~~~\text{Positive crossing}  && \text{Negative crossing}~~~~
\end{array}
\end{equation}
If instead a defect with monodromy $\color{nBLUE}a$ passes through a codimension-one transition function $\color{nORANGE}b$, which can be understood as the topological operator implementing the symmetry action, and orientation $\pm$, we have that the monodromy changes by ${\color{nBLUE}b}\longrightarrow{\color{nBLUE}b^{\mp1}ab^{\pm 1}}$, see Figure \ref{fig.transDEF}. In a nutshell, every time our defect crosses another (or passes through a transition function), the monodromy changes to a new element in the same conjugacy class.

\begin{figure}[htb!]
    \centering
    \resizebox{0.6\textwidth}{!}{%
    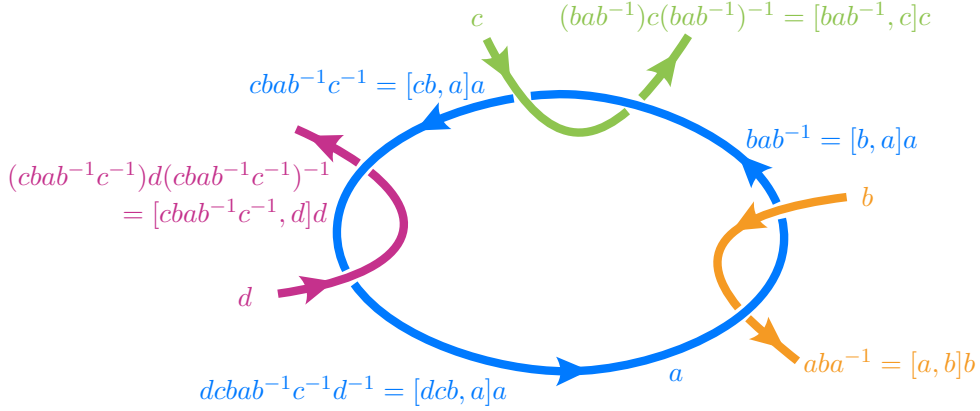
    }
    \caption{Illustration of how the linking of defects changes the monodromy around them, while staying in the same conjugacy class. For the defect associated to a monodromy $a$ (in blue) the closing condition \eqref{eq. closing comm} needs to be satisfied in order for the configuration to be consistent.}
    \label{fig.linkSHOW}
\end{figure}

Consider some defect with monodromy $a$, crossing an ordered series of defects implementing the monodromies $(d_{a,1},d_{a,2},\dots,d_{a,n_a})$. Without loss of generality we assume that the crossings are positive, since otherwise we can just take the inverse monodromy, i.e., flip the orientation of the $d_{a,i}$-defect. We also allow for transition functions, simply noting that the signs work in the opposite way. Overall, we find that the $a$-defect changes to a different representative of the conjugacy class given by
\begin{equation}
    a\to d_{a,1}ad_{a,1}^{-1}\to \dots\to(d_{a,n_a}\dots d_{a,1})a (d_{a,n_a}\dots d_{a,1})^{-1}\,.
\end{equation}
For the defect to close it needs to come back to the original element and we therefore require
\begin{equation}\label{eq. closing comm}
    [d_{a,n_a}\dots d_{a,2}d_{a,1},a]=1 \;.
\end{equation}
The same argument holds for every other defect in the brane configuration described by the link (note that for each crossing the precise element of the conjugacy class enters). We illustrate this in Figure \ref{fig.linkSHOW}.

Given some linking configuration involving $N$ defects with monodromies in the conjugacy classes $\{[x_i]\}_{i=1}^N$ in $G$, the following system of equations needs to have a solution in order to describe a well-defined configuration 
    \begin{equation}\label{eq.linkarr}
        \big\{[d_{x_i,n_i}\dots d_{x_i,2}d_{x_i,1},x_i]=1\big\}_{i=1}^N\,.
    \end{equation}
Furthermore, if the brane configuration in question is the defect needed to trivialize some class in $H_2(BG;\mathbb{Z})$ represented by $[a_1,b_1]\dots[a_g,b_g]$, by filling, e.g., the 1-cycles associated to $\{a_i\}_{i=1}^g$, additional restrictions need to be included. Let $\{y_{a_i,j}\}_j$ the monodromies of the defects (including sign accounting for their orientation) filling the 1-cycle associated to $a_i$, then
\begin{equation}\label{eq.linkarrFILL}
    \prod_jy_{a_i,j}=a_i\quad \text{for }\;i=1,\dots,g \,.
\end{equation}
Equations \eqref{eq.linkarr} and \eqref{eq.linkarrFILL}, functions of a set of $N$ unknown variables $\{x_i\}_{i=1}^N$, define a $N+g$ relations in $R$.
Unless $R$ is ``dense'' enough, in general the system of equations will be over-determined, and no linking arrangement will be possible.

        \begin{table}[hbt]
    \centering
    
    \newcolumntype{C}[1]{>{\centering\arraybackslash}m{#1}}
    \begin{tabular}{|C{0.8cm}|C{0.3\textwidth}|C{8.5cm}|}
        \hline
        Case & Links & Link conditions \\ \hline \hline
        (I) & \raisebox{-.5\height}{\centering
     \resizebox{0.3\textwidth}{!}{%
\begingroup%
  \makeatletter%
  \providecommand\color[2][]{%
    \errmessage{(Inkscape) Color is used for the text in Inkscape, but the package 'color.sty' is not loaded}%
    \renewcommand\color[2][]{}%
  }%
  \providecommand\transparent[1]{%
    \errmessage{(Inkscape) Transparency is used (non-zero) for the text in Inkscape, but the package 'transparent.sty' is not loaded}%
    \renewcommand\transparent[1]{}%
  }%
  \providecommand\rotatebox[2]{#2}%
  \newcommand*\fsize{\dimexpr\f@size pt\relax}%
  \newcommand*\lineheight[1]{\fontsize{\fsize}{#1\fsize}\selectfont}%
  \ifx\svgwidth\undefined%
    \setlength{\unitlength}{123.63764966bp}%
    \ifx\svgscale\undefined%
      \relax%
    \else%
      \setlength{\unitlength}{\unitlength * \real{\svgscale}}%
    \fi%
  \else%
    \setlength{\unitlength}{\svgwidth}%
  \fi%
  \global\let\svgwidth\undefined%
  \global\let\svgscale\undefined%
  \makeatother%
  \begin{picture}(1,0.35678522)%
    \lineheight{1}%
    \setlength\tabcolsep{0pt}%
    \put(0,0){\includegraphics[width=\unitlength,page=1]{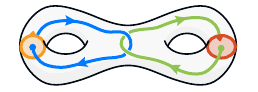}}%
    \put(0.38683295,0.02912986){\color[rgb]{0,0.47843137,1}\makebox(0,0)[lt]{\lineheight{1.25}\smash{\begin{tabular}[t]{l}$\color{nBLUE}{a}$\end{tabular}}}}%
    \put(0.02138476,0.16952854){\color[rgb]{0,0.47843137,1}\makebox(0,0)[lt]{\lineheight{1.25}\smash{\begin{tabular}[t]{l}$\color{nORANGE}{b}$\end{tabular}}}}%
    \put(0.52681988,0.29474741){\color[rgb]{0,0.47843137,1}\makebox(0,0)[lt]{\lineheight{1.25}\smash{\begin{tabular}[t]{l}$\color{nGREEN}{c}$\end{tabular}}}}%
    \put(0.93882134,0.16238736){\color[rgb]{0,0.47843137,1}\makebox(0,0)[lt]{\lineheight{1.25}\smash{\begin{tabular}[t]{l}$\color{nRED}{d}$\end{tabular}}}}%
  \end{picture}%
\endgroup%

    }} & $\begin{aligned}[c] 
[(d^{-1}cd)b^{-1},a] &= 1\\
[(b^{-1}ab)d^{-1},c]&= 1
\end{aligned}$\\ \hline
        (II) & \raisebox{-.5\height}{\centering
     \resizebox{0.3\textwidth}{!}{%
\begingroup%
  \makeatletter%
  \providecommand\color[2][]{%
    \errmessage{(Inkscape) Color is used for the text in Inkscape, but the package 'color.sty' is not loaded}%
    \renewcommand\color[2][]{}%
  }%
  \providecommand\transparent[1]{%
    \errmessage{(Inkscape) Transparency is used (non-zero) for the text in Inkscape, but the package 'transparent.sty' is not loaded}%
    \renewcommand\transparent[1]{}%
  }%
  \providecommand\rotatebox[2]{#2}%
  \newcommand*\fsize{\dimexpr\f@size pt\relax}%
  \newcommand*\lineheight[1]{\fontsize{\fsize}{#1\fsize}\selectfont}%
  \ifx\svgwidth\undefined%
    \setlength{\unitlength}{123.63764966bp}%
    \ifx\svgscale\undefined%
      \relax%
    \else%
      \setlength{\unitlength}{\unitlength * \real{\svgscale}}%
    \fi%
  \else%
    \setlength{\unitlength}{\svgwidth}%
  \fi%
  \global\let\svgwidth\undefined%
  \global\let\svgscale\undefined%
  \makeatother%
  \begin{picture}(1,0.35678522)%
    \lineheight{1}%
    \setlength\tabcolsep{0pt}%
    \put(0,0){\includegraphics[width=\unitlength,page=1]{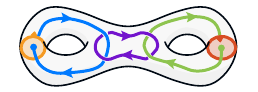}}%
    \put(0.93734531,0.1562085){\color[rgb]{0,0.47843137,1}\makebox(0,0)[lt]{\lineheight{1.25}\smash{\begin{tabular}[t]{l}$\color{nRED}{d}$\end{tabular}}}}%
    \put(0.46442737,0.04201557){\color[rgb]{0,0.47843137,1}\makebox(0,0)[lt]{\lineheight{1.25}\smash{\begin{tabular}[t]{l}$\color{nPURPLE}{x}$\end{tabular}}}}%
    \put(0.57581841,0.3143547){\color[rgb]{0,0.47843137,1}\makebox(0,0)[lt]{\lineheight{1.25}\smash{\begin{tabular}[t]{l}$\color{nGREEN}{c}$\end{tabular}}}}%
    \put(0.35954532,0.01592285){\color[rgb]{0,0.47843137,1}\makebox(0,0)[lt]{\lineheight{1.25}\smash{\begin{tabular}[t]{l}$\color{nBLUE}{a}$\end{tabular}}}}%
    \put(0.01572852,0.16606777){\color[rgb]{0,0.47843137,1}\makebox(0,0)[lt]{\lineheight{1.25}\smash{\begin{tabular}[t]{l}$\color{nORANGE}{b}$\end{tabular}}}}%
  \end{picture}%
\endgroup%

    }} & $\begin{aligned}[c] 
[xb^{-1},a] &= 1\\
[(b^{-1}abxb^{-1}a^{-1}b)d^{-1},c]&= 1\\
[(d^{-1}cd)(b^{-1}ab),x]&=1
\end{aligned}$ \\ \hline
        (III) & \raisebox{-.5\height}{\centering
     \resizebox{0.3\textwidth}{!}{%
\begingroup%
  \makeatletter%
  \providecommand\color[2][]{%
    \errmessage{(Inkscape) Color is used for the text in Inkscape, but the package 'color.sty' is not loaded}%
    \renewcommand\color[2][]{}%
  }%
  \providecommand\transparent[1]{%
    \errmessage{(Inkscape) Transparency is used (non-zero) for the text in Inkscape, but the package 'transparent.sty' is not loaded}%
    \renewcommand\transparent[1]{}%
  }%
  \providecommand\rotatebox[2]{#2}%
  \newcommand*\fsize{\dimexpr\f@size pt\relax}%
  \newcommand*\lineheight[1]{\fontsize{\fsize}{#1\fsize}\selectfont}%
  \ifx\svgwidth\undefined%
    \setlength{\unitlength}{120.68766917bp}%
    \ifx\svgscale\undefined%
      \relax%
    \else%
      \setlength{\unitlength}{\unitlength * \real{\svgscale}}%
    \fi%
  \else%
    \setlength{\unitlength}{\svgwidth}%
  \fi%
  \global\let\svgwidth\undefined%
  \global\let\svgscale\undefined%
  \makeatother%
  \begin{picture}(1,0.36550623)%
    \lineheight{1}%
    \setlength\tabcolsep{0pt}%
    \put(0,0){\includegraphics[width=\unitlength,page=1]{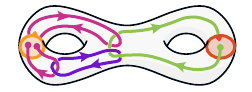}}%
    \put(0.95060007,0.16170738){\color[rgb]{0,0.47843137,1}\makebox(0,0)[lt]{\lineheight{1.25}\smash{\begin{tabular}[t]{l}$\color{nRED}{d}$\end{tabular}}}}%
    \put(0.37336497,0.00478535){\color[rgb]{0,0.47843137,1}\makebox(0,0)[lt]{\lineheight{1.25}\smash{\begin{tabular}[t]{l}$\color{nPURPLE}{x}$\end{tabular}}}}%
    \put(0.50399184,0.29642872){\color[rgb]{0,0.47843137,1}\makebox(0,0)[lt]{\lineheight{1.25}\smash{\begin{tabular}[t]{l}$\color{nGREEN}{c}$\end{tabular}}}}%
    \put(0.23681659,0.17884965){\color[rgb]{0,0.47843137,1}\makebox(0,0)[lt]{\lineheight{1.25}\smash{\begin{tabular}[t]{l}$\color{nPINK}{y}$\end{tabular}}}}%
    \put(0.00945793,0.16668541){\color[rgb]{0,0.47843137,1}\makebox(0,0)[lt]{\lineheight{1.25}\smash{\begin{tabular}[t]{l}$\color{nORANGE}{b}$\end{tabular}}}}%
  \end{picture}%
\endgroup%

    }} & $\begin{aligned}[c] 
[x,byb^{-1}] &= a\\
[(d^{-1}cd)(byb^{-1}),x]&= 1\\
[(b^{-1}xbyb^{-1}x^{-1}b)(byb^{-1}xby^{-1}b^{-1})d^{-1},c]&=1\\
[(byb^{-1}xby^{-1}b^{-1}d^{-1}cdby^{-1}bx^{-1}byb^{-1})b^{-1}xb,y]&=1
\end{aligned}$\\ \hline
        (IV) & \raisebox{-.5\height}{\centering
     \resizebox{0.3\textwidth}{!}{%
\begingroup%
  \makeatletter%
  \providecommand\color[2][]{%
    \errmessage{(Inkscape) Color is used for the text in Inkscape, but the package 'color.sty' is not loaded}%
    \renewcommand\color[2][]{}%
  }%
  \providecommand\transparent[1]{%
    \errmessage{(Inkscape) Transparency is used (non-zero) for the text in Inkscape, but the package 'transparent.sty' is not loaded}%
    \renewcommand\transparent[1]{}%
  }%
  \providecommand\rotatebox[2]{#2}%
  \newcommand*\fsize{\dimexpr\f@size pt\relax}%
  \newcommand*\lineheight[1]{\fontsize{\fsize}{#1\fsize}\selectfont}%
  \ifx\svgwidth\undefined%
    \setlength{\unitlength}{123.63764966bp}%
    \ifx\svgscale\undefined%
      \relax%
    \else%
      \setlength{\unitlength}{\unitlength * \real{\svgscale}}%
    \fi%
  \else%
    \setlength{\unitlength}{\svgwidth}%
  \fi%
  \global\let\svgwidth\undefined%
  \global\let\svgscale\undefined%
  \makeatother%
  \begin{picture}(1,0.35678522)%
    \lineheight{1}%
    \setlength\tabcolsep{0pt}%
    \put(0,0){\includegraphics[width=\unitlength,page=1]{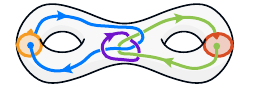}}%
    \put(0.92276016,0.17279076){\color[rgb]{0,0.47843137,1}\makebox(0,0)[lt]{\lineheight{1.25}\smash{\begin{tabular}[t]{l}$\color{nRED}{d}$\end{tabular}}}}%
    \put(0.45812931,0.0461412){\color[rgb]{0,0.47843137,1}\makebox(0,0)[lt]{\lineheight{1.25}\smash{\begin{tabular}[t]{l}$\color{nPURPLE}{x}$\end{tabular}}}}%
    \put(0.54762344,0.31150166){\color[rgb]{0,0.47843137,1}\makebox(0,0)[lt]{\lineheight{1.25}\smash{\begin{tabular}[t]{l}$\color{nGREEN}{c}$\end{tabular}}}}%
    \put(0.30684142,0.00598856){\color[rgb]{0,0.47843137,1}\makebox(0,0)[lt]{\lineheight{1.25}\smash{\begin{tabular}[t]{l}$\color{nBLUE}{a}$\end{tabular}}}}%
    \put(0.00623948,0.1678583){\color[rgb]{0,0.47843137,1}\makebox(0,0)[lt]{\lineheight{1.25}\smash{\begin{tabular}[t]{l}$\color{nORANGE}{b}$\end{tabular}}}}%
  \end{picture}%
\endgroup%

    }} & $\begin{aligned}[c] 
[x(d^{-1}c^{-1}dxd^{-1}cd)^{-1}b^{-1},a] &= 1\\
[(bab^{-1})(x^{-1}ax)^{-1}d,c]&= 1\\
[(d^{-1}cd)(b^{-1}a^{-1}bcb^{-1}ab)^{-1},x]&=1
\end{aligned}$\\ \hline
    \end{tabular}
    \caption{Examples of possible linking defect configurations for a given $G$ bundle on the genus-two Riemann surface $\Sigma_2$ associated to the homology class represented by $[a,b][c,d]$. Note that, for example, case (I) does not involve additional defects other than those associated to classes $a$ and $c$, while case (IV) is a Borromean link, where no pair of the three defects is linked. We additionally depict the (non-trivial) linking conditions \eqref{eq.linkarr} and \eqref{eq.linkarrFILL}.}
    \label{tab:linkingArrangements}
\end{table}

For intuition, in Table~\ref{tab:linkingArrangements} we depict some examples of candidate links acting as potential symmetry-breaking defects for a non-trivial element in $H_2(BG;\mathbb{Z})$ associated to $[a,b][c,d]$ and described by a $G$ bundle on a genus-two Riemann surface. In these cases we spell out the consistency conditions \eqref{eq.linkarr} explicitly.

Regardless of the existence of the internal linking configurations resolving the bouquet of defects, the defect networks associated to filling the two different components of $S^3 \backslash \Sigma_g$ will be linked with each other. In particular one finds the linking of $a_i$- and $b_i$-defects, in generalization of Figure~\ref{fig:torusfilling}.

One can further ask the question, if one can resolve the bouquet of defects with a single $g$-valent junction to a simpler brane configuration. For instance, minimal building blocks might be all possible trivalent junctions between the codimension-two defects, which can resolve the $g$-valent junction in a multitude of ways.\footnote{Note that the growth of possible resolutions grows super-exponentially (in analogy to the number of Feynman diagrams for fixed asymptotic states).} While we do not have an argument why such a resolution might be required, we will see that it occurs naturally in some string theory examples in Section~\ref{sec:juncIIA}.

\subsection{Examples}

The linking conditions \eqref{eq.linkarr} and \eqref{eq.linkarrFILL} might seem too generic to explicitly show whether a solution exists, since the elements $d_{I,j}$ will be specific representatives of the conjugacy classes of known $\{a_i,b_i\}_{i=1}^g$ and unknown monodromies $\{x_j\}_{j}$, and one can take arbitrarily complicated arrangements. In this subsection we will go through some explicit examples with groups $G$ where the existence or absence of a solution can be settled.

\subsubsection*{Surface groups}
We first consider the genus $g\geq 2$ surface group $\pi_1(\Sigma_g)$ \eqref{eq.surface group}, defined by a single relation
\begin{equation}
    \pi_1(\Sigma_g) = \langle a_1, b_1, \dots, a_g, b_g | [a_1,b_1]  \dots [a_g,b_g] \rangle \,,
\end{equation}
which, as explained around \eqref{eq.surfacegroupH2}, has $H_2(\pi_1(\Sigma_g);\mathbb{Z})\simeq \mathbb{Z}$, generated by $[a_1,b_1]  \dots [a_g,b_g]$. Now, in trying to solve \eqref{eq.linkarr} and \eqref{eq.linkarrFILL}, an important property of surface groups with genus $g\geq 2$ comes in handy, namely that the centralizer of $y$ is given by \cite[Section 1.1.3]{farb2012primer}
\begin{equation}
    Z_G(y)=\{x\in G:[y,x]=1\}\simeq\mathbb{Z}\simeq\langle r_G(y)\rangle\,,
\end{equation}
where $r_G(y)$ is the \emph{unique primitive root of $y$ in $G$}, i.e, the element such that there is an $n\geq 1$ with $r_G(y)^n=y$, and no other $w\in G$ such that $w^m=r_G(y)$ for $m>1$. For $a_i$ or $b_i$, generators of $G\simeq \pi_1(\Sigma_g)$, this implies that they are their own primitive roots. This has the nice implication that the system \eqref{eq.linkarr} can be rewritten as
\begin{equation}\label{eq. syssurf2}
     \{d_{x_j,n_j}\dots d_{x_j,1}=r_G(x_j)^{\alpha_j}\}_{j=1}^{N}\,,\quad \text{with }\, \alpha_j\in\mathbb{Z}\,.
\end{equation}
Take now the group homomorphism mapping the group to its Abelianization
\begin{equation}
    \begin{array}{lccc}
        \text{Ab}:& \quad G \simeq \pi_1(\Sigma_g)&\longrightarrow& \text{Ab} (G) \simeq\mathbb{Z}^{2g}  \\
        & x& \longmapsto&\hat{x}
    \end{array}\;,
\end{equation}
where the generators of $\text{Ab}(G)$ given by $\{\hat{a}_i,\hat{b}_i\}_{i=1}^{g}$ can be understood as an orthonormal basis of $\mathbb{R}^{2g}\supset\mathbb{Z}^{2g}$. The system of consistency conditions \eqref{eq. syssurf2} and \eqref{eq.linkarrFILL} then maps to the linear system of equations 
\begin{equation}
    \left\{\sum_{l=1}^{n_j}\hat{d}_{x_j,l}=\eta_j\hat{x}_j\right\}_{j=1}^{N}\cup\bigg\{\sum_j\hat{y}_{a_i,j}=\hat{a}_i\bigg\}_{i=1}^g\,,
\end{equation}
where $\eta_j\in\mathbb{Q}$ are some rational numbers such that $x_j^{\eta_j}=r_G(x_j)^{\alpha_j}$ (note that upon Abelianization we switch to an additive notation). We then have that the above is a system of $N+g$ equations and $N$ variables, thus being \emph{over-determined} and without solution. We thus conclude that it is not possible to resolve the bouquet of defects needed to trivialize the classes $\Omega_2^\xi(B\pi_1(\Sigma_g))$ through links, and consequently junctions between defects must be present for $g\geq 2$.
    
As an illustration, if the links in Table \ref{tab:linkingArrangements} were associated to $G\simeq\pi_1(\Sigma_2)$, then the linking conditions would map to the following linear systems upon Abelianization:
\begin{equation}
         {\rm (I)}\left\{\begin{array}{rl}
               \hat{c}-\hat{b}  &=\alpha\hat{a}  \\
               \hat{a}-\hat{d}&=\gamma\hat{c} 
            \end{array}\right.\;,\;
            {\rm (II)}\left\{\begin{array}{rl}
               \hat{x}-\hat{b}  &=\alpha\hat{a}  \\
               \hat{x}-\hat{d}&=\delta\hat{c} \\
               \hat{c}+\hat{a}&=\eta\hat{x}
            \end{array}\right.\;,\;
            {\rm (III)}\left\{\begin{array}{rl}
               0&=\hat{a}  \\
               \hat{c}+\hat{y}&=\eta\hat{x} \\
               -\hat{d}+\hat{x}+\hat{y}&=\gamma\hat{c}\\
               \hat{c}+\hat{x}&=\iota\hat{y}
            \end{array}\right.\;,\;
            {\rm (IV)}\left\{\begin{array}{rl}
               \hat{b}  &=\alpha\hat{a}  \\
               \hat{d}&=\gamma\hat{c} \\
               1&=1
            \end{array}\right.\,,
\end{equation}
none of which have solutions, since $\{\hat{a}, \hat{b},\hat{c},\hat d\}$ is an orthogonal base in $\mathbb{R}^4$. Note that for (III) one of the equations impose that $\hat{a}=0$, i.e., $a$ is a product of commutators. Note that this is not possible in surface groups, since this would imply an additional relation different than the closing one.

\subsubsection*{Groups with torsional commutators}

In general, one can argue that, given some non-trivial class in $H_2(BG;\mathbb Z)$ generated by $[a_1,b_1]\dots[a_g,b_g]$ (where, assuming minimality, $[a_i,b_i]\neq 1$ for all $i$), there can be no symmetry-breaking defect described only in terms of a link, if for some $i\in\{1,\dots g\}$ their commutator $[a_i,b_i]$ is not a torsional element, i.e., $[a_i,b_i]^n\neq 1$ for all $n\neq 0$. In order to see this, consider the handle of the Riemann surface associated to $[a_i,b_i]\neq 1$, in such a way that defects going around it implement $a_i$, see \eqref{eq.linkarrFILL}, and go through the transition function $b_i$. Now, since
\begin{equation}\label{eq. arg1}
    1\neq[a_i,b_i]=\Big[\prod_jy_{a_i,j},b_i\Big]=\Big(\prod_jy_{a_i,j}\Big)\Big(\prod_j[b_i,y_{a_i,j}]y_{a_i,j}\Big)^{-1} \,,
\end{equation}
this means that at least some defect $y_{a_{i},j_0}$ must not commute with $b_i$ (and only crosses $b_i$ once, since said set of defects are associated to $a_i$). However, this implies that the centralizer $Z_G(y_{a_{i},j_0})$ does not have $b_i$ as one of its generators, and thus there must be some additional $b_i^{\pm 1}$ factor in the closing equation \eqref{eq.linkarr}. Since $y_{a_{i},j_0}$ does not cross the transition function a second time, said term must come from an additional linking with a defect in the conjugacy class of $b_i$. In order for such linking not to be ``undone'' by simply moving the defects around, the $b_i$ defect must itself be linked with the $y_{a_{i},j_0}$. However, again because $y_{a_{i},j_0}\notin Z_G(b_i)$, an additional linking with $y_{a_{i},j_0}$ in the closing relation \eqref{eq.linkarr} for $b_i$ is needed! Every time there is a crossing the associated monodromies change in the form
\begin{equation}\label{eq. arg2}
    b_i\to [y_{a_{i},j_0},b_i]b_i\,,\quad y_{a_{i},j_0}\to[b_i,y_{a_{i},j_0}]y_{a_{i},j_0}\,,
\end{equation}
where we have omitted additional conjugating elements. If $[b_i,y_{a_{i},j_0}]$ is free, no matter how many times we cross $b_i$ and $y_{a_{i},j_0}$ (while keeping them linked), the closing relations will not be satisfied. However, if $[b_i,y_{a_{i},j_0}]^n=1$ for some $n$, then after $n$ mutual crossings we might be able to recover the initial $y_{a_{i},j_0}$ and $b_i$, which could allow for the defects to close.

We can illustrate this with a simple example given by 
\begin{equation}\label{eq. group pow}
    G=\big\langle a,b\mid [a,b]^g,\big[[a,b] ,a\big],\big[[a,b] ,b\big]\big\rangle\,,
\end{equation}
i.e, the individual generators are not torsion, but $[a,b]$ is, and further commutes with all elements. Its first two homology groups are given by
\begin{equation}
    H_1(BG;\mathbb{Z})\simeq\mathbb{Z}\oplus\mathbb{Z}\,,\qquad H_2(BG;\mathbb{Z})\simeq\mathbb{Z}\oplus\mathbb{Z}_g\oplus\mathbb{Z}_g\,,
\end{equation}
with the individual factors of the second homology generated by $[a,b]^g$, $\big[[a,b],a\big]$ and $\big[[a,b],b\big]$. While the torsional generators can be represented by a genus-one surface, $[a,b]^g$ is associated to a genus-$g$ Riemann manifold where all handles have identical transition functions. Using 
\begin{equation}
    (yxy^{-1})[y,x^k]y^{-1}(yxy^{-1})^{-1}=[y,x^{k+1}]y^{-1}\,,
\end{equation}
and\footnote{To show this, we proceed by induction. Notice that
\begin{equation}
    [x^2,y]=\big[x,[x,y]\big][x,y]^2\,,
\end{equation}
so that $[x^2,y]=[x,y]^2\mod [F',F]$. Now, assuming $[x^k,y]=[x,y]^k\mod [F,F']$, we have
\begin{equation}
    [x^{k+1},y]=[x^kx,y]=\big[x^k,[x,y]\big][x,y][x,y]^k=[x,y]^{k+1}\mod [F,F']\,.
\end{equation}
} 
\begin{equation}
    [x^k,y]=[x,y]^k\mod [F,F']\,,\quad\text{with }F'=[F,F]\,,
\end{equation}
(where for our explicit group \eqref{eq. group pow} is an exact equality, since $F'$ is central) one can construct explicit arrangement of linking defects that act as a symmetry-breaking defect in the interior of $\Sigma_g$ without the need of junctions. We illustrate said arrangement in Figure \ref{fig.OnlyLinks}. While this group might seem a little \emph{ad hoc}, in Appendix \ref{app:Mth} we will describe an explicit M-theory construction that realizes \eqref{eq. group pow} for $g=2$.

\begin{figure}[ht]
\begin{center}
\begin{subfigure}[b]{0.5\textwidth}
\centering
     \resizebox{0.8\textwidth}{!}{%
    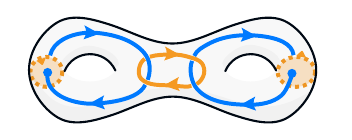
    }
\caption{Linking arrangement for $g=2$.} \label{ff.OL2}
\end{subfigure}
\begin{subfigure}[b]{0.49\textwidth}
\centering
     \resizebox{0.8\textwidth}{!}{%
    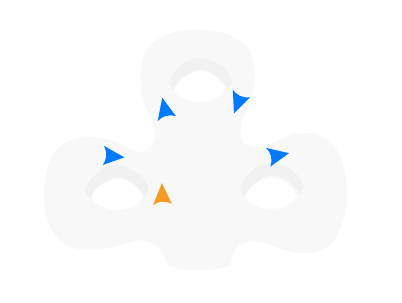
    }
\caption{Linking arrangement for generic $g>2$.} \label{ff.OLg}
\end{subfigure}
\caption{Linking configurations of defects trivializing the class in $\Omega_2^\xi(BG)$ represented by the generator of the second homology group of \eqref{eq. group pow}. Note that in showing that such arrangement can be realized without junctions, we have used that for our group \eqref{eq. group pow} $[b,a^g]=[b,a]^g=1$ and $([b,a^k]b^{-1})bab^{-1}([b,a^k]b^{-1})^{-1}=a$.
\label{fig.OnlyLinks}}
\end{center}
\end{figure}
We conclude this section by noting that, even though the generator of $H_2(BG;\mathbb{Z})$ can be represented by $[a,b]^g$, it can be shown, see \cite[Example 2.6]{CULLER1981133}, that for any group $G$ and $u,\,v\in G$, $[u,v]^k$ can be expressed as the product of $\big\lfloor\frac{k}{2}\big\rfloor+1$ commutators (generally different ones). There should then be another representative of $[a,b]^g$ described by a $G$ bundle in a Riemann manifold of genus $\big\lfloor\frac{g}{2}\big\rfloor+1$, though it is not obvious that the transition functions that of the $\Sigma_{\lfloor g/2\rfloor+1}$ $G$ bundle would admit an arrangement of defects that does not involve junctions.


\section{Junctions of axion strings in type IIA}
\label{sec:juncIIA}

In this section we will discuss a more involved top-down example resulting in a monodromy group $G$ where some generators of $H_2(G;\mathbb{Z})$ require the product of two commutators and junctions between codimension-two defects are needed to trivialize the associated bordism classes. For this we consider type IIA string theory compactified on a Calabi-Yau threefold $X$, resulting in a $\mathcal{N}=2$ theory in four dimensions. We will focus on the hypermultiplet sector, which enjoys a monodromy group generically given by the form \cite{Alexandrov:2010np, Alexandrov:2011va,Alexandrov:2013yva}
\begin{equation}
    \label{eq:monogroupIIA}
    \Gamma_{\rm hyper}= \mathrm{SL}(2;\mathbb{Z})\ltimes\big(\Gamma_{\rm mirror}\ltimes {\rm H}_{2h^{2,1}+3}(\mathbb{Z})\big)\,,
\end{equation}
where $\Gamma_{\rm mirror} \subset \mathrm{Sp}(2h^{1,1}+2;\mathbb{Z})$ is the associated monodromy group of the mirror Calabi-Yau $X^\vee$ (since the hypermultiplet moduli space $\mathcal{M}_{\rm h}$ is fibered over the complex structure moduli space $\mathcal{M}_{\rm cs,mirror}$ of $X^\vee$) and ${\rm H}_{2h^{2,1}+3}(\mathbb{Z})$ is the Heisenberg group with $2h^{2,1}+2$ generators (over the integers). We will focus on the Heisenberg group, since it will have interesting properties for our purposes. While the full structure in \eqref{eq:monogroupIIA} might influence the bordism group due to the action of the semi-direct products, here, we focus on its Heisenberg subgroup. It would be interesting to extend our analysis to also include transition functions in SL$(2;\mathbb{Z})$ and $\Gamma_{\text{mirror}}$.

Let for our purposes $H_3(X;\mathbb Z)$, the middle homology of the internal space, be free of torsion, and consider the associated symplectic base $\{\alpha^I,\beta_I\}_{I=0}^{h^{2,1}}$, in such a way that, reducing the 3-form $C_3$ with respect to this basis, one obtains a number of $\{\chi_I,\tilde{\chi}^I\}_{I=0}^{h^{2,1}}$ axions in the lower-dimensional effective theory. Additionally, one has the universal axion $\sigma$ originating from compactifying the 6-form $\tilde{B}_6$ dual to the NSNS 2-form $B_2$ on the entire $X$ (or equivalently dualizing the unreduced $B_2$ field in four dimensions). In principle the $(2h^{2,1}+3)$ scalars would have independent axionic shift symmetries, however, the topological term of the 10d (massless) type IIA action,
\begin{equation}\label{eq.CSIIA}
    2\kappa_{10}^2S_{\rm IIA}\supset-\frac{1}{2}\int B_2\wedge\dd C_3\wedge\dd C_3 \,,
\end{equation}
reduces to 
\begin{equation}
    2\kappa_{10}^2S_{\rm 4d}\supset-\frac{1}{2}\int B_2\wedge\dd \chi_I\wedge\dd \tilde \chi^I=\frac{1}{4}\int\dd B_2\wedge\big(\dd \chi_I\wedge \tilde \chi^I- \chi_I\wedge\dd \tilde \chi^I\big)\,.
\end{equation}
This topological term requires a non-trivial mixing between the different axionic symmetries,
\begin{equation}\label{eq.axions IIA}
    \left\{\begin{array}{l}
    \chi_I \to \chi_I+\zeta_I \,, \\
    \tilde \chi^I \to \tilde \chi^I+\tilde \zeta^I \,, \\
    ~\,\sigma\to\sigma+\alpha-\frac{1}{2}(\zeta_I\tilde \chi^I-\tilde\zeta^I \chi_I) \,,
    \end{array}\right. \quad \text{for }I=0,\dots,h^{2,1}\,,
\end{equation}
which precisely corresponds to the Heisenberg algebra of ${\rm H}_{2h^{2,1}+3}$, with presentation
\begin{equation}\label{eq.HEIS}
    {\rm H}_{2h^{2,1}+3}(\mathbb{Z})=\big\langle a_0,b_0,\dots,a_{h^{2,1}},b_{h^{2,1}},z\mid [z,a_I],[z,b_I],[a_I,a_J],[b_I,b_J],[a_I,b_J]z^{-\delta_{IJ}}\big\rangle\,,
\end{equation}
where $a_I$ and $b_I$ are the shifts associated to $\chi_I$ and $\tilde \chi^I$ and $z$ corresponds to the shift of $\sigma$. Noticing that, since, for example, $z=[a_0,b_0]$, we can simplify the presentation by removing $z$ from the set of relations and it is straightforward to compute that (see also \cite{LEE1996230})
\begin{subequations}
    \begin{align}
        H_1\big(B{\rm H}_{2h^{2,1}+3}(\mathbb{Z});\mathbb{Z}\big)&=\mathbb{Z}^{2h^{2,1}+2}\,,\\ H_2\big(B{\rm H}_{2h^{2,1}+3}(\mathbb{Z});\mathbb{Z}\big)&=\left\{\begin{array}{ll}
                 \mathbb{Z}^2&  \text{for }h^{2,1}=0 \,, \\
                \mathbb{Z}^{2h^{2,1}(h^{2,1}+1)}\oplus\mathbb{Z}^{h^{2,1}} &\text{for }h^{2,1}\geq 1 \,.
            \end{array}\right. \label{eq:homologyHeisen}
    \end{align}
\end{subequations}
The second homology group strongly depends on $h^{2,1}$. In the special case of vanishing $h^{2,1}$ we can write 
\begin{equation}
    {\rm H}_3(\mathbb{Z})\simeq\langle a,b\mid \big[a,[a,b]\big],\,\big[b,[a,b]\big]\rangle \,,
\end{equation}
and the non-trivial elements of the second homology group are precisely associated to $\big[a,[a,b]\big]$ and $\big[b,[a,b]\big]$ via Hopf's theorem. Both are therefore described in terms of a single commutator. On the other hand, for non-vanishing $h^{2,1}$ the elements in $H_2\big(B\mathrm{H}_{2h^{2,1}}(\mathbb{Z}); \mathbb{Z}\big)$ fall into two classes
\begin{itemize}
    \item $[a_I,a_J] \,, [b_I,b_J] \,,[a_I,b_J]$ with $I \neq J$: They form 
    \begin{equation}
        \binom{h^{2,1}+1}{2}+\binom{h^{2,1}+1}{2}+ h^{2,1}(h^{2,1}+1)=2h^{2,1}(h^{2,1}+1) \,,
    \end{equation}
    independent elements in $R \cap [F,F]$, which are not in $[R,F]$ and generate $\mathbb{Z}^{2h^{2,1}(h^{2,1}+1)}$ in \eqref{eq:homologyHeisen}. All of them can be written in terms of a single generator.
    \item $[a_I,b_I][a_0,b_0]^{-1}$ with $I \neq0$: They are obtained by using $z=[a_0,b_0]$ to eliminate $z$, and again are elements of $R \cap [F,F]$ which are not in $[R,F]$, generating the remaining $\mathbb{Z}^{h^{2,1}}$ in \eqref{eq:homologyHeisen}. However, these need to be written in terms of two commutators!
\end{itemize}

Note that since
\begin{equation}
    [z,a_I]=\big[[a_0,b_0],a_I\big]=\big[a_0,[b_0,a_I]\big]\big[[a_0,a_I],b_0\big]\quad \text{(analogously for $[z,b_I]$)}\,,
\end{equation}
we have that $[z,a_I],\,[z,b_I]\in[R,F]$ and thus are not associated to non-trivial elements in $H_2\big({\rm H}_{2h^{2,1}+3}(\mathbb{Z});\mathbb{Z}\big)$.

Summarizing, for $h^{2,1}\geq 1$ we find that $H_2\big(B{\rm H}_{2h^{2,1}+3}(\mathbb{Z});\mathbb{Z}\big)$ has $h^{2,1}$ generators that need to be expressed as the product of (at least) two commutators. Using the arguments in Section~\ref{ss.HigherGenusAndLinking} the trivialization of the associated bordism class requires a $S^1\vee S^1$ arrangement of defects that realize the appropriate monodromies of the genus-two surface representing $[a_I,b_I][a_0,b_0]^{-1}$. Moreover, as argued around \eqref{eq. arg1} and \eqref{eq. arg2}, since the elements $\{a_I,b_I\}_{I=0}^{h^{2,1}}$ are free, and the central element $z=[a_I,b_I]$ is not torsion (i.e., $z^k\neq 1$ for any $k\neq 0$), there is no possible resolution of the $S^1\vee S^1$ arrangement of defects in the form of linking defects and junctions are required.

\begin{figure}[htb]
    \centering
    \resizebox{0.75\textwidth}{!}{%
    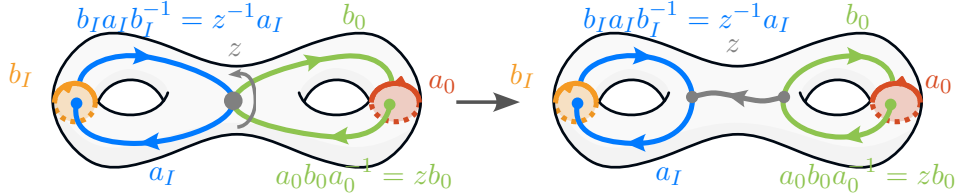
    }
    \caption{Sketch of the defect network needed to kill the bordism class generated by $[a_I,b_I][a_0,b_0]^{-1}$. The defects correspond to D4-branes wrapping an appropriate 3-cycle class in $H_3(X;\mathbb{Z})$. We also depict the resolution of the four-valent junction into two trivalent ones involving the defect with monodromy $z$, which in our setup is a fundamental string.}
    \label{fig.D4res}
\end{figure}

These can be realized in different ways, as depicted in Figure~\ref{fig.D4res}. On one hand, we can have a single four-valent junction where the defects with monodromies $a_I^{-1}$, $z^{-1}a_I$, $zb_0$ and $b_0^{-1}$ meet (since the product of said monodromies, in any order, is the trivial element, the junction is consistent). On the other hand, we could resolve said four-valent junction into two trivalent ones, joined by a defect, with monodromy $z$ for the resolution shown in Figure \ref{fig.D4res}. One could wonder whether this resolution is possible, and whether the defect with monodromy $z$ exists, since it did neither corresponds to a non-trivial element in $H_1\big(B{\rm H}_{2h^{2,1}+3}(\mathbb{Z});\mathbb{Z}\big)$ nor $H_2\big(B{\rm H}_{2h^{2,1}+3}(\mathbb{Z});\mathbb{Z}\big)$. However, for the type IIA example it is obvious that indeed this is the case!

Remember that the $a_I$ monodromy is associated to the shift of the $\chi_I$ axion obtained from reducing $C_3$ on the 3-cycle $\alpha^I$. Since a D4-brane is magnetically charged under $C_3$, the defect implementing $a_I$ in four dimensions will be a D4-brane wrapped on the Poincar\'e-dual 3-cycle $\beta_I$, where $\alpha^I \cdot \beta_I=1$. Analogous arguments imply that $b_I$ is implemented by a D4-brane wrapped on $\alpha^I$. The $z$ monodromy is instead implemented directly by the universal axion $\sigma$ originating from the NSNS sector and its defect is given by an unwrapped fundamental type IIA string. Thus, the resolution of the defect network into two trivalent junctions is realized by a fundamental string stretched between two D4-branes wrapping different internal 3-cycles.

This demonstrates how for this case the Cobordism Conjecture implies the existence of junctions between wrapped D4-branes, which can be resolved through a fundamental string stretching between them. The existence of the topological term \eqref{eq.CSIIA} was crucial for our derivation, since it induces a non-trivial mixing in the shift symmetries of RR and NSNS axions. An analogous argument would follow in the mirror-symmetric type IIB setup, where the associated topological term, rewritten in the democratic formulation \cite{Bergshoeff:2001pv} is
\begin{equation}
    2\kappa_{10}^2 S_{\rm IIB}\supset-\frac{1}{2}\int C_4\wedge\dd B_2\wedge\dd C_2=\frac{1}{2}\int B_2\wedge\big( \dd C_2\wedge\dd C_4-\dd C_0\wedge\dd C_6\big)\,.
\end{equation}
By reducing $C_2$ and $-C_4+\frac{1}{2}B_2\wedge C_2$ on the Poincar\'e-dual basis of 2- and 4-cycles $\{\alpha_I,\Sigma^I\}_{I=1}^{h^{1,1}}$, as well as $-C_6+B_2\wedge C_4-\frac{1}{3}B_2\wedge B_2\wedge C_2$ on the entire $X$, we obtain a set of axions $\{\chi_I,\tilde\chi^I\}_{I=0}^{h^{1,1}}$, see \cite{Alexandrov:2008gh} for more details. Now the defects implementing the shifts would be D5- and D3-branes wrapping the dual 4- and 2-cycles, together with a D7-brane wrapping the whole Calabi-Yau (for the $C_0$ shift) and the unwrapped D1-string (for the $C_6$ axion). The central element $z$ of the Heisenberg group is again implemented by the fundamental string, which can stretch between the axion string originating from D1-, D3-, D5- and D7-branes resolving the four-valent junctions. 

We expect that this type of non-linear topological mixing between different fields reducing to axions upon compactification to always result in Heisenberg-like groups or generalizations thereof. In Appendix~\ref{app:Mth} we show how the reduction of the topological term $2\kappa_{11}^2S_{\rm 11d}\supset-\frac{1}{6}\int C_3\wedge\dd C_3\wedge \dd C_3$ in M-theory results in generalized Heisenberg groups with several central elements. This opens the door to the construction of finite symmetry groups with generators of $H_2(BG;\mathbb{Z})$ associated to the product of a large number of commutators and potential resolutions by linking brane networks without the need for junctions. See also \cite{Berasaluce-Gonzalez:2012abm,Grimm:2015ona,Corvilain:2016kwe} for how said generalized Heisenberg groups appear in string theory, e.g., type IIB orientifold compactifications to $\mathcal{N}=1$ supergravity theories in four dimensions as well as their F-theory lift (including torsion elements in the group relations through compactification on torsion cycles).

Finally, we would expect that similar relations could arise in heterotic string theory, where the Bianchi identity is modified to $\dd H_3=\frac{\alpha'}{4}\big({\rm tr}\mathcal{R}^2-{\rm tr}F^2\big)$, and there is the Green-Schwarz term in the action $S_{\rm het}\supset\int B_2\wedge{\rm tr}F^4$. One would then expect, to find generalized Heisenberg-like relations involving the NSNS axion and the gauge field degrees of freedom after compactification on a suitable 6-manifold.


\section{Conclusion and Outlook}
\label{sec:concl}

In this paper we systematically study the symmetry-breaking defects associated to $\Omega^{\xi}_2 (BG)$ for discrete symmetry groups $G$. The discrete nature of the symmetry group naturally suggests the involvement of codimension-two defects, which are classified by the monodromy around them given by an element in $G$. By using Hopf's theorem for group homology, we are able to relate the non-trivial bordism classes related to $H_2(BG;\mathbb{Z})$ to $G$ bundles on genus-$g$ Riemann surfaces $\Sigma_g$. This also resolves the mismatch in dimensionality since not the individual branes are the obstructions to a deformation to nothing but their non-trivial configurations.

If the non-trivial element in $H_2(BG;\mathbb{Z})$ can be associated to a single commutator, the non-trivial deformation class is given by a 2-torus $T^2$ with commuting $G$ transition function around the basis of closed paths. In this case the bordism defect is the circle compactification of a codimension-two defect which passes through another $G$ transition function. This can be interpreted as the non-trivial linking of the two codimension-two defects. For more than one generators one requires higher-genus Riemann surfaces to detect the non-trivial deformation classes, in which case the bordism defects themselves contain junctions (or linking in special cases). 

Thus, rather then predicting the appearance of genuinely codimension-three objects the classes in $\Omega^{\xi}_2 (BG)$, can be trivialized by non-trivial properties of codimension-two objects including the existence of brane networks!

We verify these predictions in Calabi-Yau compactifications of type II theories resulting in $\mathcal{N}=2$ supergravity in four dimensions, for which $G$ is given by a Heisenberg group with non-trivial $H_2(BG;\mathbb{Z})$. We find examples involving genus-two surfaces that predict non-trivial string junctions. In the full string theory these are described nicely by branes ending on branes induced by the non-trivial Bianchi identities of the RR-fields.

It might be tempting to expect that similar networks are also relevant for defects associated to $\Omega^{\xi}_k (BG)$, with $k > 2$. However, there are two obstacles for such a generalization. First, it is not clear how the junctions or linking can be associated to objects of even higher codimension, since these configurations are associated to the intersection of a codimension-two defect with a codimension-one transition function, i.e., a configuration in codimension three. Second, once one goes to higher bordism groups, the differentials are generically non-trivial and while $H_k (BG;\mathbb{Z})$ appears on the second page of the Atiyah-Hirzebruch spectral sequence it might not survive to its $\infty$-page. Instead one could imagine a generalization of the brane systems in this work which involves lower-dimensional objects, which naturally carry features in higher codimension. Evidence of this was seen for example in \cite{Debray:2023yrs, Braeger:2025kra}, where some bordism generators were fibrations, which might allow for an interpretation in terms of brane networks (see also \cite{Nevoa:2025xiq}).

The most interesting configurations in our investigation have all been associated to Heisenberg groups and generalizations thereof. These originate from the compactification of theories (here; high-dimensional supergravity theories) whose action contains a topological term among the $p$-form fields of the parent theory. The topology of the internal space is then imprinted via an intersection matrix in the topological couplings of the low-energy effective theories. It is a natural question to ask, whether the groups we encounter here are the most general or different types of compactification manifolds might lead to different classes, potentially with different origin than the topological terms in the action.

The connection of the brane configurations describing the symmetry-breaking defects and the topological terms and non-trivial Bianchi identities of the UV theory also suggests the connection to higher-groups (see, e.g., \cite{Cordova:2018cvg, Benini:2018reh} for an introduction for physicists). Indeed, since higher-groups naturally mix various $p$-form symmetries, they are a promising arena to explore the interplay between their dual (magnetic) defects, their intersections and brane networks.

\section*{Acknowledgements}

We are grateful to Damian van de Heisteeg, Miguel Montero, Ethan Torres, and Irene Valenzuela for useful discussions and comments. We especially thank Ethan Torres for comments on the manuscript. The work of the authors is supported by the European Union through ERC Starting Grant SymQuaG-101163591 StG-2024.


\begin{appendix}

\section{Heisenberg groups from M-theory}
\label{app:Mth}

In this appendix we will generalize the compactifications of type IIA string theory on a Calabi-Yau 3-fold described in Section~\ref{sec:juncIIA} to the more general setting of M-theory on various 7-manifolds. As we will see, the extra dimension on which we compactify will result in more than one 2-forms in the four-dimensional theory, which results in generalized Heisenberg groups describing the axion shift symmetries. We start by considering the topological term in the eleven-dimensional $\mathcal{N}=1$ supergravity action, which lifts \eqref{eq.CSIIA} to
\begin{equation}\label{eq. CSMth}
    2\kappa_{11}^2S_{\rm 11d}\supset-\frac{1}{6}\int C_3\wedge\dd C_3\wedge \dd C_3\,.
\end{equation}
Compactifying our theory on a generic manifold $Y_7$ allows us to expand the 3-form $C_3$ as 
\begin{equation}
    C_3=B^i_2\wedge\eta_i+\chi^I\omega_I\,,\quad\text{with }\left\{\begin{array}{l}
         H^1(Y_7;\mathbb{Z})\simeq\langle\{\eta_i\}_{i=1}^{b_1}\rangle \,, \\
         H^3(Y_7;\mathbb{Z})\simeq\langle\{\omega_I\}_{I=1}^{b_3}\rangle \,,
    \end{array}\right.
\end{equation}
where for the sake of simplicity we will not consider torsion for the time being and only focus on the expansion in 1- and 3-forms giving rise to axions $\{\chi^I\}_{I=1}^{b_3}$ and 2-forms $\{B^i_2\}_{i=1}^{b_1}$ in four dimensions. From \eqref{eq. CSMth} we obtain
\begin{equation}
    2\kappa_{11}^2 S_{\rm 4d}\supset \frac{\mathsf{c}_{iJK}}{2}\int B^i\wedge\dd \chi^I\wedge\dd \chi^J=-\frac{\mathsf{c}_{iJK}}{4}\int{\dd B^i}\wedge\big(\dd \chi^J\wedge \chi^K-\chi^J\wedge\dd \chi^K\big)\,,
\end{equation}
where we define the intersection matrix
\begin{equation}
    \mathsf{c}_{iJK}=\int_{Y_7}\eta_i\wedge\omega_I\wedge\omega_J\,,\qquad\text{with }\mathsf{c}_{iJK}=-\mathsf{c}_{iKJ}\,.
\end{equation}
Additionally, considering the set of axions $\{\sigma^i\}_{i=1}^{b_1}$ dual to the 2-forms $B_2^i$ (equivalently resulting from compactifying the dual $\tilde{C}_6$ on the 6-cycle classes from $H_6(Y;\mathbb{Z})\simeq H^1(Y;\mathbb{Z})$), this translates to a transformation law for the various axions, given by
\begin{equation}\label{eq. axions M-theory}
    \left\{\begin{array}{l}
            \chi^I\to \chi^I+\zeta^I \,,\\
            \sigma^i\to\sigma^i+\alpha^i+\frac{1}{2}\mathsf{c}_{iJK}\chi^J\zeta^K \,,
    \end{array}\right.
\end{equation}
which generalizes the type IIA axion shift symmetries in \eqref{eq.axions IIA}. Taking $\{x_I\}_{I=1}^{b_3}$ and $\{z_i\}_{i=1}^{b_1}$ to be the generators of the shift symmetries acting on the $\chi^I$ and $\sigma^i$ axions, respectively, this results in the algebraic relations
\begin{equation}\label{eq. pres gen}
    [x_J,x_K]=\prod_{i=1}^{b_1}z_i^{\mathsf{c}_{iJK}} \,,
\end{equation}
which generalizes the Heisenberg group \eqref{eq.HEIS} to
\begin{equation}
    G=\Big\langle x_1,\dots,x_{b_3},z_1,\dots,z_{b_1}\Big|[z_i,x_J],[z_i,z_j],[x_I,x_J]\big(\prod_iz_i^{\mathsf{c}_{iJK}}\big)^{-1}\Big\rangle\,.
\end{equation}
Note that while the intersection matrix $\mathsf{c}_{iJK}$ is non-degenerate (i.e., has full rank), its inverse $\mathsf{w}^{bJK}$, with $\mathsf{c}_{aJK}\mathsf{w}^{bJK}=\delta_a^b$, might not be an integer matrix. We can define the following linear map\footnote{Remember that the elements $\hat{x}_I$ and  $\hat{z}_i$ denote the image of $x_I$ and $z_i$ in Ab$(G)$.}
\begin{equation}
    \begin{array}{lccc}\label{eq. c func}
         \mathsf{c}:&\Lambda^2(\mathbb{Z}^{b_3})&\longrightarrow&\mathbb{Z}^{b_1}  \\
         & \hat{x}_I\wedge\hat{x}_J&\longmapsto&\sum_{i=1}^{b_1}\mathsf{c}_{iJK}\hat{z}_i
    \end{array}\,,
\end{equation}
which, even if it might have full rank, might not be surjective. Upon Abelianization we have that \eqref{eq. pres gen} translates to $\sum_i \mathsf{c}_{iJK}\hat z_i=0$. This way one obtains
\begin{equation}\label{eq.ghom1}
    H_{1}(G;\mathbb{Z})= \text{Ab}(G) = \mathbb{Z}^{b_3}\oplus\frac{\mathbb{Z}^{b_1}}{{\rm Im}(\mathsf{c})}\,.
\end{equation}
We now turn our attention to the second homology group of $G$. Similar to the traditional Heisenberg group, the commutators $[z_i,x_J]$ and $[z_i,z_j]$ are non-trivial elements in $(R\cap[F,F])/[R,F]$, thus generating some free parts $\mathbb{Z}^{b_1 b_3}$ and $\mathbb{Z}^{\frac{1}{2}b_1(b_1-1)}$ in $H_2(G;\mathbb{Z})$, with generators given in terms of a single commutator. Writing now 
\begin{equation}\label{eq. def wa}
    w_{\mathbf{a}}=\prod_{1\leq J< K\leq b_3}[x_J,x_K]^{a_{JK}}=\prod_{1\leq J< K\leq b_3}\Big(\prod_iz_i^{\mathsf{c}_{iJK}}\Big)^{a_{JK}}=\prod_iz_i^{\sum_{J< K}\mathsf{c}_{iJK}a_{JK}}\,,
\end{equation}
in order for $w_{\mathbf{a}}\in R\cap[F,F]$ we need 
\begin{equation}
    \sum_{1\leq J< K\leq b_3}\mathsf{c}_{iJK}a_{JK}=0\quad\forall\,i=1,\dots,b_1\,,
\end{equation}
which is precisely given by the kernel of \eqref{eq. c func}, $\langle\{w_{\mathbf{a}}\}_{\mathbf{a}}\rangle\simeq \ker(\mathsf{c})\simeq \mathbb{Z}^{\frac{1}{2}b_3(b_3-1)-{\rm rank}(\mathsf{c})}$. Being clear that the elements in \eqref{eq. def wa} are not in $[R,F]$, we thus conclude that 
\begin{equation}\label{eq.ghom2}
    H_2(G;\mathbb{Z})\simeq \mathbb{Z}^{b_1b_3}\oplus\mathbb{Z}^{\frac{1}{2}b_1(b_1-1)}\oplus \mathbb{Z}^{\frac{1}{2}b_3(b_3-1)-{\rm rank}(\mathsf{c})}\,.
\end{equation}
The generators of the first two factors can be written as single commutators, $[z_i,x_J]$ and $[z_i,z_j]$. As for $w_{\mathbf{a}}=\prod_{1\leq J< K\leq b_3}[x_J,x_K]^{a_{JK}}$ generators of the last factor, we see that they are represented by skew-symmetric matrices in $\Lambda^2(\mathbb{Z}^{b_3})$, whose rank is bounded by $2\lfloor\frac{b_3}{2}\rfloor$. We thus have that the minimal number of commutators the generators of $H_2(G;\mathbb{Z})$ will be expressed as will be at most $1+\big\lfloor\frac{b_3}{2}\big\rfloor$. In principle, one could expect that, choosing $Y_7$ with $b_3$ large enough, one could engineer generators of $H_2(G;\mathbb{Z})$ that needed a large number of commutators. However, we will see that in general things are not as straightforward.
\begin{itemize}
    \item We first consider $Y_7=T^7$, compactifying down to $\mathcal{N}=8$ supergravity in four dimensions, with $b_1=\binom71=7$ and $b_3=\binom73=35$. We identify each 1-cycle by $i=1,\dots ,7$ and 3-cycle by $I\in\{[ijk]\,,i,j,k=1,\dots,7\}$, so that
    \begin{equation}\label{eq.intT7}
        \mathsf{c}_{iJK}=\mathsf{c}_{i[jkl][mnp]}=\int_{T^7}\dd x^i\wedge(\dd x^j\wedge\dd x^k\wedge\dd x^l)\wedge(\dd x^m\wedge\dd x^n\wedge\dd x^p)=\epsilon_{ijklmnp}\,,
    \end{equation}
    with $\epsilon_{1234567}=1$. Since $\mathsf{c}:\Lambda^2(\mathbb{Z}^{35})\to\mathbb{Z}^{7}$ is clearly surjective, we obtain
    \begin{equation}
        H_1(G;\mathbb{Z})\simeq \mathbb{Z}^{35}\,,\qquad H_2(G;\mathbb{Z})\simeq \mathbb{Z}^{245}\oplus\mathbb{Z}^{21}\oplus\mathbb{Z}^{588}\,.
    \end{equation}
    In order to compute the number of commutators of the generators of the $\mathbb{Z}^{588}$ factor, we see from \eqref{eq.intT7} that $\mathsf{c}_{i[jkl][mnp]}=0$ if any index is repeated. This implies that only commutators $[x_{[jkl]},x_{[mnp]}]$ with non-overlapping 3-cycles are non-trivial, amounting to a total of $\binom76\times\frac{1}{2}\binom63=70$ possibilities, and it is clear that a given commutator in \eqref{eq. pres gen} equals to a single $z_i^{\pm 1}$. Since we can write
    \begin{equation}
        z_1=[x_{[234]},x_{[567]}]\,,\quad z_2=[x_{[134]},x_{[567]}]\,,\quad\text{etc}.\,,
    \end{equation}
    we conclude that every of said 70 possibilities can be written as the product of two commutators. This was already the case in the Heisenberg groups that appeared in type IIA Calabi-Yau compactifications, so there is no surprise here.
    \item We next consider $Y_7=T^3\times{\rm K3}$, down to a $\mathcal{N}=4$ supergravity in four dimensions, with $b_1=3$ and $b_3=1+3\times22=67$. Note that 1-cycles come from the torus, and 3-cycles consist in either the whole 3-torus or a torus 1-cycle times a K3 2-cycle $\{[\omega_\alpha]\}_{\alpha=1}^{22}$. With this in mind, it is clear that the intersection matrix is 0 if the 3-cycle is the $T^3$, while for the other case,
    \begin{equation}
        \mathsf{c}_{iJK}=\mathsf{c}_{i(j\alpha)(k\beta)}=\int_{T^3\times{\rm K3}}\dd x^i\wedge\dd x^j\wedge\omega_\alpha\wedge\dd x^k\wedge\omega_\beta=\epsilon_{ijk}\eta_{\alpha\beta}\,,
    \end{equation}
    where $\eta_{\alpha\beta}$ is the intersection matrix of K3, symmetric and with signature $(3,19)$, since $H^2({\rm K3};\mathbb{Z})$ is isomorphic to $2(-E_8) \oplus 3U$). Since the determinant of $\eta_{\alpha\beta}=\pm 1$, the image of $\mathsf{c}:\Lambda^2(\mathbb{Z}^{67})\rightarrow\mathbb{Z}^{3}$ is $\mathbb{Z}^3$, and one recovers
    \begin{equation}
        H_1(G;\mathbb{Z})\simeq \mathbb{Z}^{67}\,,\qquad H_2(G;\mathbb{Z})\simeq \mathbb{Z}^{201}\oplus\mathbb{Z}^3\oplus\mathbb{Z}^{2208}\,.
    \end{equation}
    As before, we will be interested in the generators of the $\mathbb{Z}^{2208}$ factor, and the minimal number of commutators they need to be expressed as. We know that when the whole 3-cycle is $T^3$ the intersection matrix vanishes (and thus the associated generators will have commutator length 1), so we can consider elements of the type $\mathsf{c}_{iJK}=\mathsf{c}_{i(j\alpha)(k\beta)}=\epsilon_{ijk}\eta_{\alpha\beta}$. Due to the Levi-Civita tensor appearing, we cannot have the same $T^3$ 1-cycle appearing more than once (and since we require 3-cycles, all of them must appear). Then from \eqref{eq. pres gen}
    \begin{equation}        [x_{(j\alpha)},x_{(k\beta)}]=\prod_{i=1}^3z_{i}^{\epsilon_{ijk}\eta_{\alpha\beta}}\,=z_{i}^{\epsilon_{ijk}\eta_{\alpha\beta}}
    \end{equation}
    with $i$ the element in $\{1,2,3\}$ left after removing $j$ and $k$. Since $|\det\eta|=1$, we can invert the relation and express $z_1$, $z_2$ and $z_3$ as single commutators. This means that, similar to the $T^7$ case, the generators of $H_2(G;\mathbb{Z})$ will be given by the product of at most two commutators.
    \item For the next example we take $Y_7=X_6\times S^1$, with $X_6$ is a Calabi-Yau 3-fold, down to a $\mathcal{N}=2$ supergravity. However, we know that $b_1(X_6)=0$, and thus $b_1(Y_7)=1$, and we are back to the case studied in Section \ref{sec:juncIIA}, consistent with type IIA string theory corresponding to M-theory on $S^1$. 
    \item We finally consider compactifying M-theory on a $G_2$ manifold down to a minimal $\mathcal{N}=1$ supergravity. However, it is known that the first Betti number for this class of manifolds is exactly zero \cite{JoyceG2}, which prevents us from considering this setting.
\end{itemize}
One might think that we have exhausted the possible M-theory settings, since the above possibilities are the only ones that preserve supersymmetry for compactifications of eleven-dimensional supergravity to four-dimensional Minkowski space, see e.g., \cite{Duff:1986hr,Acharya:2004qe}. However, that would be jumping to a conclusion to fast. One important assumption in the above derivation was $Y_7$ not having torsion in $H_1(Y_7;\mathbb{Z})$ and $H_3(Y_7;\mathbb{Z})$. We will see that relaxing this requirement will allow us to obtain a genus-two generator of $H_2(G;\mathbb{Z})$ whose defects do not need junctions, since they admit a linking resolution. Additionally, if one allows for (not too drastic) supersymmetry breaking, we see that more freedom for the realization of the intersection matrix $\mathsf{c}_{iJK}$ is possible.

\subsubsection*{An example with links: 
M-theory on $G_2$-manifold with torsion}

When discussing the compactification of M-theory on a $G_2$-manifold above, we swiftly concluded that it was not possible to get a $\int B_2\wedge\dd\chi\wedge\dd\tilde\chi$ term for the four-dimensional action as there was no 1-cycle on which to reduce $C_3\to B_2$, since $b_1=0$, \cite{JoyceG2}. However, one could still find a torsion part 
\begin{equation}
    H_1(Y_7;\mathbb{Z})={\rm Ab}\big(\pi_1(Y_7)\big)\simeq\bigoplus_{i=1}^N\mathbb{Z}_{p_i}\,,
\end{equation}
if $Y_7$ is not simply-connected.\footnote{See \cite[Section 12.4]{joyce2000compact} for Joyce manifolds (a particular set of $G_2$ manifolds given by ${T}^7$ quotients) where $H_1(Y_7;\mathbb{Z})\simeq\mathbb{Z}_2^a$.} As an illustrative example, we follow \cite{Kachru:2001je,Camara:2011jg,Kaufmann:2026tsy}, and consider a $G_2$-manifold with a type IIA orbifold limit, given by 
\begin{equation}
    Y_7=\frac{(X_6\times S^1)/\hat{\sigma}}{\hat\tau}\,,\quad \text{with }\left\{\begin{array}{l}
          \hat{\sigma}(x,y)=\big(\sigma(x),-y\big) \,,\\
          \hat\tau(x,y)=\big(\tau(y),y+\frac{1}{2}\big) \,,
    \end{array}\right.
\end{equation}
with $X_6$ a Calabi-Yau 3-fold and where $\sigma$ and $\tau$ are, respectively, commuting orbifold and freely acting involutions, with $\sigma(\tau(x))=\tau(\sigma(x))$. Through this construction the obtained $G_2$-manifold has $H_1(Y_7;\mathbb{Z})\simeq\mathbb{Z}_2$.
 
We now consider two different (free) 4-cycles in $Y_7$ built by 
\begin{equation}
    \Sigma_A=S_4\times\{0,\tfrac{1}{2}\}\,,\quad \Sigma_B=S_3\times S^1\,\,,
\end{equation}
where $S_4$ and $S_3$ are respectively a 4-cycle and 3-cycle in $X_6$. In order to have $\Sigma_A$ and $\Sigma_B$ invariant under the two involutions, $S_4$ and $S_3$ must be respectively even and odd under $\sigma$ and $\tau$. Intersecting both cycles on the Calabi-Yau results in a one-dimensional curve $C_1=S_4\cap S_3\subset X_6$, which is odd under $\sigma$. Then $\tilde\gamma=C_1\times\{0,\frac{1}{2}\}$ lifts to a $\mathbb{Z}_2$-torsional 1-cycle  which precisely corresponds to the intersection $\Sigma_A\cap\Sigma_B=\gamma$, even if the 4-cycles were free. This translates in a intersection $\mathsf{c}_{\gamma AB}=1\mod 2$.

We now compactify the topological term \eqref{eq. CSMth}, which results in two axions $\chi$ and $\tilde\chi$ coming from the reduction of $C_3=\chi[\Sigma_A]+\tilde\chi[\Sigma_B]+\dots$ (with $[\Sigma_A]$ and $[\Sigma_B]$ the $H^3(Y_7;\mathbb{Z})$ cohomology classes Poincar\'e dual to $\Sigma_A$ and $\Sigma_B$). Additionally, we have a $\mathbb{Z}_2$-valued phase $\sigma\in\{0,\frac{1}{2}\}$ dual to the $\mathbb{Z}_2$ 2-form in four dimensions originating from reducing $C_3$ on $\gamma$ (equivalently from the expansion of the dual 6-form as $C_6=\sigma[\gamma]+\dots$). Taking $a$, $b$ and $z$ to be the associated generators of the shifts on $\chi$, $\tilde\chi$ and $\sigma$, we find the relations
\begin{equation}
    [z,a]=[z,b]=1\,,\quad z^2=1\,,\quad [a,b]=z\,,
\end{equation}
within the full duality group $G=F/R$. Barring fine-tuned additional relations coming from the remainder of the $Y_7$ geometry, the above generates a (sub)group such as the one described in \eqref{eq. group pow} for $g=2$, which translates to $[a,b]^2$ being one of the $H_2(G;\mathbb{Z})$ generators. As discussed there in this case one can find a defect network without junctions, since the four-valent junction can be resolved by a linking configurations of strings in four dimensions coming from the wrapping of M5-branes on $\Sigma_A$ and $\Sigma_B$, see Figure \ref{ff.OL2}.

\subsubsection*{More than two commutators: 
M-theory on a twisted 7-torus}

In the above discussion we have seen how compactifications of M-theory on manifolds that resulted in a four dimensional theory with $\mathcal{N}\geq1$ supersymmetry always resulted in intersection matrices $\mathsf{c}_{iJK}$ being too simple, in such a way that generators of $H_2(G;\mathbb{Z})$ were always expressed by at most two commutators. We will see that, if we allow for $Y_7$ to completely break supersymmetry, there is more room to find longer commutator products. For this let us consider a twisted 7-torus $X_7$, see e.g. \cite{Hull:2005hk}, only ``slightly'' breaking supersymmetry completely in four dimensions, spanned by 1-forms $\{e^i\}_{i=1}^7$ satisfying the Maurer-Cartan equation
\begin{equation}
    \dd e^a=-\frac{1}{2}f^{a}_{bc} e^b\wedge e^c\,,
\end{equation}
where the structure constants $f^{a}_{bc}$ are understood as \emph{metric fluxes}. It is clear that then the first Betti number is given by $b_1=7-{\rm rank}(f^a_{bc})$. We thus define a basis $\{\eta_i=A_{ia}e^a\}_{i=1}^{b_1}$. As for the 3-forms $\omega=\frac{1}{6}\Omega_{abc}e^a\wedge e^b\wedge e^c$, the Betti number $b_3$ will be given by the number of independent solutions to
\begin{equation}
    \dd\omega=0\Rightarrow \Omega_{a[bc} f^a_{de]} =0\,,\qquad\omega\neq \dd\beta\;\text{with }\,\beta=\tfrac{1}{2}B_{ij}e^i\wedge e^j\Rightarrow \Omega_{dec} \neq -3 B_{b[c} f^b_{de]}\,,
\end{equation}
for any $B_{ij}$.\footnote{In practice, in order to compute $H^k(X;\mathbb{Z})$ it is more straightforward to simply recursively take the set of closed $k$-forms (taking into account the Maurer-Cartan equation) and mod out by the image of all possible $(k-1)$-forms under $\dd_{k-1}:\Lambda^{k-1}(X)\to \Lambda^k(X)$.} We thus take $\{\omega_I=\frac{1}{6}\Omega_{I,abc}e^a\wedge e^b\wedge e^c\}_{I=1}^{b_3}$ as the generator set of such solutions. It is then clear that the intersection matrix would be given by
\begin{equation}
    \mathsf{c}_{iJK}=\int_{X_7}\eta_i\wedge\omega_I\wedge\omega_J=\frac{1}{36}\epsilon^{abcdefg} A_{ia}\Omega_{I,bcd}\Omega_{J,efg}\,,
\end{equation}
from which our usual analysis follows. We illustrate this for a particular example, with Maurer-Cartan equations
\begin{equation}
    \dd e^1=\dd e^2=\dd e^3=\dd e^7=0\,,\quad\dd e^4=e^1\wedge e^2\,,\quad \dd e^5=e^2\wedge e ^3\,,\quad \dd e^6=e^3\wedge e^1\,,
\end{equation}
so that the only non-trivial structure constants are $f^4_{12}=f^5_{23}=f^6_{31}=-2$. We have that clearly $b_3=4$, with the basis of 1-forms being $\{e^1,e^2,e^3,e^7\}$. As for the 3-forms, one iteratively computes that $b_3=20$, with 
\begin{equation}
    \begin{array}{c}
        \omega_1=e^{367},\;\omega_2=e^{357},\;\omega_3=e^{356} ,\;\omega_4=e^{347} ,\;\omega_5=-e^{156}+e^{346},\\\omega_6=e^{256}+e^{345} ,\;\omega_7=-e^{157}+e^{267} ,\;
        \omega_8=e^{257},\;\omega_9=e^{247},\;\omega_{10}=-e^{145}+e^{246} ,\\\omega_{11}=e^{245} ,\;\omega_{12}=e^{236},\;\omega_{13}=e^{235} ,\;\omega_{14}=e^{234} ,
        \;
        \omega_{15}=e^{167},\\\omega_{16}=e^{147},\;\omega_{17}=e^{146} ,\;\omega_{18}=e^{136} ,\;\omega_{19}=e^{124},\;\omega_{20}=e^{235} ,
    \end{array}
\end{equation}
where $e^{abc}=e^a\wedge e^b\wedge e^c$. Then one can obtain the intersection matrix $\mathsf{c}_{iJK}$, in such a way that the only non-trivial commutators are\footnote{We do not explicitly write the intersection matrix $\mathsf{c}_{iJK}$, since it would require four $20\times 20$ skewsymmetric matrices. However, these are quite sparse, and the non-zero entries can be read from the commutators in \eqref{eq.twistedCOMM}.} 
\begin{subequations}\label{eq.twistedCOMM}
        \begin{align}
            [x_{20},x_3]=[x_5,x_{14}]=[x_6,x_{19}]=[x_{10},x_{12}]=[x_{11},x_{18}]=[x_{17},x_{13}]=z_1\label{eq.comm1}\\
[x_{10},x_{4}]=[x_5,x_{9}]=[x_6,x_{16}]=[x_{17},x_{8}]=[x_{11},x_{15}]=z_2
        \\
             [x_1,x_{10}]=[x_2,x_{17}]=[x_{16},x_{3}]=[x_7,x_{5}]=[x_{15},x_{6}]=z_3,\\
             [x_1,x_{11}]=[x_{10},x_{2}]=[x_3,x_{9}]=[x_4,x_{6}]=[x_8,x_{5}]=[x_6,x_{7}]=z_7\label{eq.comm4}\\             
             [x_4,x_{5}]=z_3^2,\quad [x_{10},x_7]=z_2^2
        \end{align}
\end{subequations}
Note that we can always write the central elements $z_1,\,z_2,\,z_3$ and $z_4$ as single commutator, so we can get rid of the $z_i$ elements and rewrite the relations $[x_J,x_K]=z_i^{\pm 1}$ in \eqref{eq.comm1} to \eqref{eq.comm4} as a generator of $H_2(G;\mathbb{Z})$ expressed by the product of two commutators. However, for $[x_4,x_{5}]=z_3^2$ and $[x_{10},x_7]=z_2^2$ this is not possible, giving rise to two generators of $H_2(G;\mathbb{Z})$ that require a product of {\it at least three  commutators}!

\begin{figure}[ht!]
    \centering
     \resizebox{\textwidth}{!}{%
    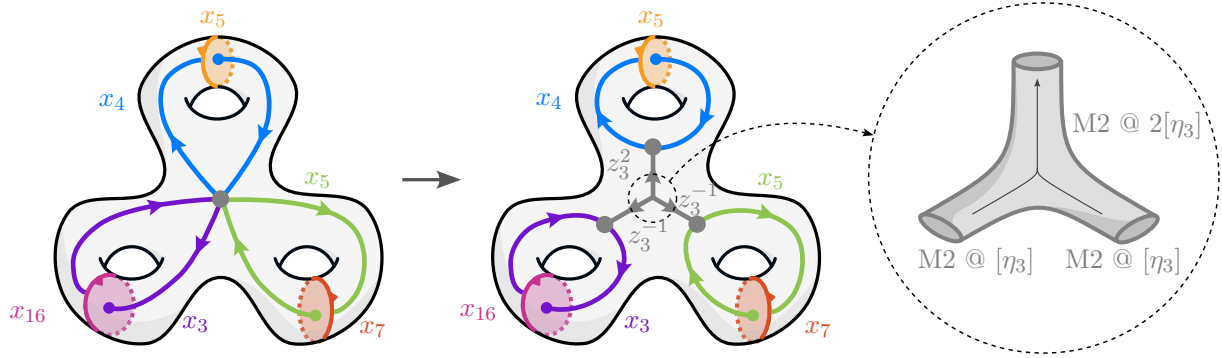
    }
    \caption{Sketch of the defect arrangement for the $[x_4,x_5][x_5,x_7][x_3,x_{16}]$ representative of a non-trivial $H_2(G;\mathbb{Z})$. The defects associated to $x_i$ correspond to M5-branes wrapped on 4-cycles, while the central $z_i$ monodromies are realized by wrapped M2-branes. Note that the six-valent junction can be resolved by junctions corresponding to M2-branes ending on M5-branes. The microscopic interpretation of the ``junction'' where $z_3^2$, $z_3^{-1}$ and $z_3^{-1}$ meet does not correspond to M2-branes ending on other M2-branes, but rather a single M2-brane wrapping a 2-chain.}
    \label{fig:threeGEN}
\end{figure}

The resulting defects would be given by three strings in four dimensions, obtained by compactifying M5-branes on the 4-cycles $\{[\omega_I]\}_{I=1}^{b_3}$ Poincar\'e dual to the 3-forms, arranged in a $\bigvee_{i=1}^3 S^1$ bouquet. The six-valent junction cannot be resolved through linking configuration of defects, since none of the commutators $[x_I,x_J]$ are torsion, as argued around \eqref{eq. arg1} and \eqref{eq. arg2}. However, it is possible to resolve the six-valent junction into four trivalent ones, involving defects implementing the $\{z_i\}_{i=1}^{b_1}$ monodromies, given by strings obtained from wrapping M2-branes on the non-trivial 1-cycles, see Figure \ref{fig:threeGEN}. Three of said junctions correspond to M2-branes ending on M5-branes, as in \cite{Strominger:1995ac}, while the fourth corresponds to a single M2-brane wrapping a 2-chain.

\end{appendix}

\bibliographystyle{JHEP}
\bibliography{ref}
\end{document}